\documentclass[preprint,prd,aps,showpacs,showkeys,nofootinbib]{revtex4}
\usepackage{graphicx}
\begin{document}
\title{Charged Lepton Flavor Violation in the B-L symmetric SSM}
\author{Xing-Xing Dong$^{1,2,3}$\footnote{dongxx@hbu.edu.cn},
Shu-Min Zhao$^{1,2,3}$\footnote{zhaosm@hbu.edu.cn},
Jia-Peng Huo$^{1,2,3}$,
Tong-Tong Wang$^{1,2,3}$\footnote{wtt961018@163.com},
Tai-Fu Feng$^{1,2,3,4}$\footnote{fengtf@hbu.edu.cn}}

\affiliation{$^1$ Department of Physics, Hebei University, Baoding, 071002, China\\
$^2$ Hebei Key Laboratory of High-precision Computation and Application of Quantum Field Theory, 071002, China\\
$^3$ Hebei Research Center of the Basic Discipline for Computational Physics, Baoding, 071002, China\\
$^4$ Department of Physics, Chongqing University, Chongqing 401331, China}

\begin{abstract}
Charged lepton flavor violation (CLFV) represents a clear new physics (NP) signal beyond the standard model (SM). In this work, we investigate CLFV processes $l_j^-\rightarrow l_i^- \gamma$ utilizing mass insertion approximation(MIA) in the minimal supersymmetric extension of the SM with local B-L gauge symmetry (B-LSSM). The MIA method can provide a set of simple analytic formulae for the form factors and the associated effective vertices, so that the movement of the CLFV decays $l_j^-\rightarrow l_i^- \gamma$ with the sensitive parameters will be intuitively analyzed. Considering the SM-like Higgs boson mass and the muon anomalous dipole moment (MDM) within $4\sigma$, $3\sigma$ and $2\sigma$ regions, we discuss the corresponding constraints on the relevant parameter space of the model.
\end{abstract}


\maketitle

\section{Introduction}
Although the Standard Model(SM) is considered as a much mature theory, it holds lepton number is conserved, that is, no charged lepton flavor violation (CLFV) occurs\cite{SMLFV}. But the CLFV processes can easily occur in new physics (NP) beyond the SM. Therefore, if the CLFV signals are observed in the future experiments, it is obvious evidence of the NP beyond the SM. In TABLE I, we show the latest experimental data for the CLFV processes $l_j^-\rightarrow l_i^- \gamma$\cite{merexp,tauerexp,taumurexp}, and these processes have been discussed in various theoretical frameworks\cite{CLFVNP1,CLFVNP2,CLFVNP3,CLFVNP4,CLFVNP5,CLFVNP6,CLFVNP7,CLFVNP8,CLFVNP9}. In this work, we investigate these CLFV processes in the minimal supersymmetric extension of the SM with local B-L gauge symmetry (B-LSSM)\cite{B-LSSM1,B-LSSM2,B-LSSM3,B-LSSM4,B-LSSM5}. We hope to reveal some properties of high energy physics through detailed analyses of these CLFV processes.
\begin{table}[h]
\caption{ \label{tab1} The latest experiment limits for the CLFV processes $l_j^-\rightarrow l_i^- \gamma$.}
\footnotesize
\begin{tabular*}{115mm}{@{\extracolsep{\fill}}ccc}
\toprule  CLFV process&Present limit&confidence level (CL) \\
\hline
$\mu\rightarrow e\gamma$ \hphantom{0} & \hphantom{0}$< 4.2\times10^{-13}$\cite{merexp}& \hphantom{0}$90\%$ \\
$\tau\rightarrow e\gamma$ \hphantom{0} & \hphantom{0}$<3.3\times10^{-8}$\cite{tauerexp}& \hphantom{0}$90\%$ \\
$\tau\rightarrow \mu\gamma$ \hphantom{0} & \hphantom{0}$<4.2\times10^{-8}$\cite{taumurexp}& \hphantom{0} $90\%$ \\
\hline
\end{tabular*}%
\end{table}

It is worth noting that we use a novel calculation method called as
mass insertion approximation(MIA)\cite{g-2MIA,htaumuMIA,HliljMIA,ZlklmMIA,Zhaog-2MIA,WangljlirMIA} to study CLFV processes $l_j^-\rightarrow l_i^- \gamma$ in the B-LSSM. The CLFV decays $l_j^-\rightarrow l_i^- \gamma$ are produced via one-loop contributions, which are influenced by the flavor mixing among the three generations of the B-LSSM sleptons and/or sneutrinos. The MIA works with the sleptons(sneutrinos) in the electroweak interaction eigenstate instead of mass eigenstate. That is to say, the MIA method operates mass insertions inside the propagators of the electroweak interaction sleptons(sneutrinos) eigenstates, instead of performing the exact diagonalization of the mass basis involved in the full one-loop computation. The MIA method has been studied by other LFV works, including the $h,H,A\rightarrow \tau\mu$ decays induced from SUSY loops\cite{htaumuMIA}, effective LFV $Hl_il_j$ vertex from right-handed neutrinos\cite{HliljMIA}, one-loop effective LFV $Zl_kl_m$ vertex from heavy neutrinos\cite{ZlklmMIA}, LFV decays $l_j\rightarrow l_i \gamma$ in the $U(1)_XSSM$\cite{WangljlirMIA} and so on. These works provide references and guidance for our research of CLFV processes $l_j^-\rightarrow l_i^- \gamma$ in the B-LSSM.

On the base of the minimum supersymmetric Standard Model (MSSM)\cite{MSSM1,MSSM2,MSSM3,MSSM4}, B-LSSM extends the gauge symmetry group to $SU(3)_C\otimes{SU(2)_L}\otimes{U(1)_Y}\otimes{U(1)_{B-L}}$, where $B$ represents the baryon number and $L$ stands for the lepton number. The B-LSSM adds two singlet Higgs superfields $\hat{\eta}$ and $\hat{\bar{\eta}}$ and three generations of right-handed neutrinos superfields $\hat{\nu}_i^c$ to the MSSM. The invariance under $U(1)_{B-L}$ gauge group imposes the R-parity conservation, which is assumed in the MSSM, to avoid proton decay\cite{B-L R Parity}. Besides, through the additional singlet Higgs states and right-handed (s)neutrinos, additional parameter space in the B-LSSM is released from the LEP, Tevatron and LHC constraints to alleviate the hierarchy problem of the MSSM\cite{B-L hierarchy1,B-L hierarchy2}. Furthermore, the B-LSSM can provide much more dark matter (DM) candidates than that in the MSSM\cite{B-LDM1,B-LDM2,B-LDM3,B-LDM4}.

Our research of CLFV processes $l_j^-\rightarrow l_i^- \gamma$ in the B-LSSM possesses much differences comparing that of $U(1)_XSSM$. Firstly, because the two models contain different fields, and the corresponding quantum numbers are different, the two works are discussed under different models. In the $U(1)_XSSM$, three Higgs singlets and right-handed neutrinos are added to MSSM. This model relieves the so called little hierarchy problem that appears in the MSSM. $\hat{S}$ is the singlet Higgs superfield with a non-zero VEV $v_s/\sqrt{2}$. The terms $\mu\hat{H}_u\hat{H}_d$ and $\lambda_H\hat{S}\hat{H}_u\hat{H}_d$ can produce an effective $\mu_{eff} =\mu + \lambda_Hv_s/\sqrt{2}$, which relieves the $\mu$ problem. Comparing with the condition in MSSM, the lightest CP-even Higgs mass at tree level is improved. The second light neutral CP-even Higgs can be at TeV order. Then it easily satisfies the constraints for heavy Higgs from experiments. Secondly, two models contain different parameters, so there are big differences in the analytical calculations, analyses at the analytical level and numerical discussions. In our work, we further discuss the numerical results changing with sensitive parameters within $4\sigma$, $3\sigma$ and $2\sigma$ specifically. We study the two-dimensional distribution of sensitive parameters under experimental constraints, and then the influences of some sensitive parameters on $Br(l_j^-\rightarrow l_i^- \gamma)$ are discussed by one-dimensional graphs. There are some differences in numerical discussion methods and ideas comparing with LFV decays $l_j\rightarrow l_i \gamma$ in the $U(1)_XSSM$.

Depended on the mass eigenstates of the particles and rotation matrixes, the mass eigenstate method is often not intuitive and clear enough to find the sensitive parameters, which will lead us to pay too much attention on many unimportant parameters. However, the MIA method provides very simple analytic formulae for the form factors, involved which can be written explicitly in terms of the sensitive parameters after a proper expansion. We can easily find the direct impacts of sensitive parameters on CLFV at the analytic level. Therefore, using MIA method to study CLFV processes will provide a new way to study other CLFV processes in the future.

The paper is organized as follows. In Sec.II, we  introduce the B-LSSM briefly including the superpotential and the general soft breaking terms. In Sec.III, we give analytic expressions for muon MDM and the CLFV ratios $Br(l_j^-\rightarrow l_i^- \gamma)$ in the B-LSSM. The numerical analyses are given in Sec.IV, and the conclusion is discussed in Sec.V. The tedious formulae are collected in Appendix A. In Appendix B, we discuss chirality flips with two examples, and demonstrate that the chirality flips occurring in the internal gaugino lines may yield dominant contributions comparing in the external lepton lines. In Appendix C, we emphasize the contributions from the incident lepton is dominant by comparing the amplitudes of FIG.\ref{figb}(a1) and (a2).
\section{the B-LSSM}
\subsection{The B-LSSM}
The B-LSSM extends the superfields of the MSSM by introducing $U(1)_{B-L}$ gauge superfield. Therefore, the local gauge group of the B-LSSM is defined as $SU(3)_C\otimes{SU(2)_L}\otimes{U(1)_Y}\otimes{U(1)_{B-L}}$. Compared with the MSSM, the B-LSSM adds two singlet Higgs superfields $\hat{\eta}$ and $\hat{\bar{\eta}}$ and three generations of right-handed neutrinos superfields $\hat{\nu}_i^c$. In the TABLE \ref{B-Lquantum numbers}, we discuss the quantum numbers of gauge symmetry group for the chiral fields in the B-LSSM. Then the B-LSSM superpotential is deduced as
\begin{table}[t]
\caption{ \label{B-Lquantum numbers}  The chiral superfields and quantum numbers in the B-LSSM.}
\footnotesize
\begin{tabular}{|c|c|c|c|c|c|}
\hline
Superfield & Spin 0 & Spin \(\frac{1}{2}\) & Generations & \(U(1)_Y\otimes\, \text{SU}(2)_L\otimes\, \text{SU}(3)_C\otimes\, U(1)_{B-L}\) \\
\hline
\(\hat{H}_d\) & \(H_d\) & \(\tilde{H}_d\) & 1 & \((-\frac{1}{2},{\bf 2},{\bf 1},0) \) \\
\(\hat{H}_u\) & \(H_u\) & \(\tilde{H}_u\) & 1 & \((\frac{1}{2},{\bf 2},{\bf 1},0) \) \\
\(\hat{Q}_i\) & \(\tilde{Q}_i\) & \(Q_i\) & 3 & \((\frac{1}{6},{\bf 2},{\bf 3},\frac{1}{6}) \) \\
\(\hat{L}_i\) & \(\tilde{L}_i\) & \(L_i\) & 3 & \((-\frac{1}{2},{\bf 2},{\bf 1},-\frac{1}{2}) \) \\
\(\hat{D}_i^c\) & \(\tilde{D}_i^c\) & \(D_i^c\) & 3 & \((\frac{1}{3},{\bf 1},{\bf \overline{3}},-\frac{1}{6}) \) \\
\(\hat{U}_i^c\) & \(\tilde{U}_i^c\) & \(U_i^c\) & 3 & \((-\frac{2}{3},{\bf 1},{\bf \overline{3}},-\frac{1}{6}) \) \\
\(\hat{E}_i^c\) & \(\tilde{E}_i^c\) & \(E_i^c\) & 3 & \((1,{\bf 1},{\bf 1},\frac{1}{2}) \) \\
\(\hat{\nu}_i^c\) & \(\tilde{\nu}_i^c\) & \(\nu_i^c\) & 3 & \((0,{\bf 1},{\bf 1},\frac{1}{2}) \) \\
\(\hat{\eta}\) & \(\eta\) & \(\tilde{\eta}\) & 1 & \((0,{\bf 1},{\bf 1},-1) \) \\
\(\hat{\bar{\eta}}\) & \(\bar{\eta}\) & \(\tilde{\bar{\eta}}\) & 1 & \((0,{\bf 1},{\bf 1},1) \) \\
\hline
\end{tabular}
\end{table}
\begin{eqnarray}
&&W_{B-L} =Y_{u,ij}\hat{Q}_i\hat{H}_u\hat{U}_j^c-Y_{d,ij}\hat{Q}_i\hat{H}_d\hat{D}_j^c +Y_{e,ij}\hat{L}_i\hat{H}_d\hat{E}_j^c+\mu_H\hat{H}_d\hat{H}_u\nonumber\\&&~~~~~~~~ +Y_{x,ij}\hat{\nu}_i^c\hat{\eta}\hat{\nu}_j^c+Y_{\nu,ij}\hat{L}_i\hat{H}_u\hat{\nu}_j^c
-{\mu_{\eta}}\hat{\eta}\hat{\bar{\eta}},
\end{eqnarray}
where $i, j$ represent the generation indices, $Y_{u,ij},Y_{d,ij},Y_{e,ij},Y_{x,ij}$ and $Y_{\nu,ij}$ correspond to the Yukawa coupling coefficients. $\mu_H$ and $\mu_{\eta}$ are both the parameters with mass dimension. $\mu_H$ indicates the supersymmetric mass between $SU(2)_{L}$ Higgs doublets $\hat{H}_d$ and $\hat{H}_u$, as well as $\mu_{\eta}$ represents the supersymmetric mass between $U(1)_{B-L}$ Higgs singlets $\hat{\eta}$ and $\hat{\bar{\eta}}$.

In the B-LSSM, the Higgs doublets and Higgs singlets obtain the nonzero vacuum expectation values, then the $SU(2)_L\otimes U(1)_Y\otimes U(1)_{B-L}$ gauge group breaks to $U(1)_{em}$.
\begin{eqnarray}
&&H_d^0 = \frac{1}{\sqrt{2}}( \phi_d + v_d  + i \sigma_d),\;\;
H_u^0 = \frac{1}{\sqrt{2}} (\phi_u  +  v_u  + i \sigma_u),
\nonumber\\&&\eta = \frac{1}{\sqrt{2}}( \phi_{\eta} + v_{\eta}  + i \sigma_{\eta} ),\;\;\;\;\;
\bar{\eta} = \frac{1}{\sqrt{2}} (\phi_{\bar{\eta}}  +  v_{\bar{\eta}}  + i \sigma_{\bar{\eta}}).
\label{4Higga}
\end{eqnarray}
Here, we define $u^2=v_{\eta}^2+v_{\bar{\eta}}^2,v^2=v_d^2+v_u^2$ and $\tan\beta'=\frac{v_{\bar{\eta}}}{v_\eta}$ in analogy to the definition $\tan\beta=\frac{v_u}{v_d}$ in the MSSM.

Correspondingly, the soft breaking terms in the B-LSSM are generally written as
\begin{eqnarray}
&&{\cal L}_{{soft}}=
- m_{\tilde{q},{i j}}^{2}\tilde{Q}_{{i}}^*\tilde{Q}_{{j}}- m_{\tilde{u},{i j}}^{2}\tilde{U}_{{i}}^*\tilde{U}_{{j}}- m_{\tilde{d},{i j}}^{2}(\tilde{D}^c_{{i}})^* \tilde{D}_{{j}}^c- m_{\tilde{L},{i j}}^{2}\tilde{L}_{{i}}^*\tilde{L}_{{j}}- m_{\tilde{E},{i j}}^{2}(\tilde{E}^c_{{i}})^* \tilde{E}_{{j}}^c\nonumber \\
&&\hspace{1.1cm}
-m_{{H}_d}^2 |{H}_d|^2-m_{{H}_u}^2 |{H}_u|^2-m_{{\eta}}^2 |{\eta}|^2-m_{{\bar\eta}}^2 |{\bar\eta}|^2
- m_{\tilde{\nu},{i j}}^{2}(\tilde{\nu}^c_{{i}})^* \tilde{\nu}_{{j}}^c+\Big[-B_{\mu}{H}_d {H}_u\nonumber \\&&\hspace{1.1cm}-B_{\eta}{\eta}{\bar\eta}
+T_{u}^{i j} \tilde{Q}_i \tilde{U}_j^c H_u+T_{d}^{i j} \tilde{Q}_i \tilde{D}_j^c H_d+T_{e}^{i j} \tilde{L}_i \tilde{E}_j^c H_u+T_{\nu}^{i j}H_u \tilde{\nu}^c_i \tilde{L}_j+T_{x}^{i j}{\eta} \tilde{\nu}^c_i\tilde{\nu}^c_j\nonumber \\
&&\hspace{1.1cm}-\frac{1}{2}({M}_1\lambda_{\tilde{B}}\lambda_{\tilde{B}} +{M}_2\lambda_{\tilde{W}}\lambda_{\tilde{W}}+{M}_3\lambda_{\tilde{g}}\lambda_{\tilde{g}}+2{M}_{B B'}\lambda_{\tilde{B}'}\lambda_{\tilde{B}}+{M}_{B'}\lambda_{\tilde{B}'}\lambda_{\tilde{B}'})+h.c.\Big],
\end{eqnarray}
where $\lambda_{\tilde{B}},\lambda_{\tilde{W}},\lambda_{\tilde{g}}$ and $\lambda_{\tilde{B}'}$ are the gauginos of $U(1)_Y, SU(2)_L, SU(3)_C$ and $U(1)_{B-L}$ respectively. Besides, the soft breaking terms of the B-LSSM include the mass squared terms of squarks, sleptons, sneutrinos and Higgs bosons, the trilinear scalar coupling terms and the Majorana mass terms.

Comparing with the MSSM or other SUSY models, the two Abelian groups in the B-LSSM produce a new effect called as the gauge kinetic mixing. Although both approaches are equivalent, it is easier to work with noncanonical covariant derivatives instead of off-diagonal field-strength tensors in practice. Hence, the covariant derivatives of the B-LSSM can be considered as
\begin{eqnarray}
D_{\mu}=\partial_{\mu}-i\left(Y, B-L\right)\left(\begin{array}{cc}
g_{Y}, & g_{Y B}^{\prime} \\
g_{B Y}^{\prime}, & g_{B-L}
\end{array}\right)\left(\begin{array}{c}
A_{\mu}^{\prime Y} \\
A_{\mu}^{\prime BL}
\end{array}\right),
\end{eqnarray}
where $Y$ and $B-L$ correspond to the hypercharge and B-L charge, as well as $A_{\mu}^{\prime Y}$ and $A_{\mu}^{\prime BL}$ denote the gauge fields of $U(1)_Y$ and $U(1)_{B-L}$. With the condition of the two Abelian gauge groups unbroken, choosing matrix R in a proper form, one can write the coupling matrix as
\begin{eqnarray}
\left(\begin{array}{cc}
g_{Y}, & g_{Y B}^{\prime} \\
g_{B Y}^{\prime}, & g_{B-L}
\end{array}\right) R^{T}=\left(\begin{array}{cc}
g_{1}, & g_{Y B} \\
0, & g_{B}
\end{array}\right),
\end{eqnarray}
here, $g_1$ corresponds to the measured hypercharge coupling which is modified in B-LSSM as given along with $g_B$ and $g_{YB}$ in Refs.\cite{gBgYBB-L}. Then, we can redefine the $U(1)$ gauge fields
\begin{eqnarray}
R\left(\begin{array}{c}
A_{\mu}^{\prime Y} \\
A_{\mu}^{\prime BL}
\end{array}\right)=\left(\begin{array}{c}
A_{\mu}^{Y} \\
A_{\mu}^{BL}
\end{array}\right).
\end{eqnarray}

The one-loop corrections for the CLFV processes are related with the mass matrices, which can be obtained by SARAH\cite{SARAH1,SARAH2}. Besides, the CLFV processes by MIA need to consider the trilinear couplings under the interaction eigenstate, so we show some couplings needed in this work as follows. The lepton-charginos-CP even(odd) sneutrinos are deduced as:
\begin{eqnarray}
&&{\cal L}_{\bar{l}_j\chi^-\tilde{\nu}^R}=\frac{i}{\sqrt{2}}\bar{l}_j
\tilde{\nu}_L^R[Y_l^jP_L\tilde{H}^-+g_2P_R\tilde{W}^-],\nonumber \\&&
{\cal L}_{\bar{l}_j\chi^-\tilde{\nu}^I}=\frac{1}{\sqrt{2}}\bar{l}_j
\tilde{\nu}_L^I[-Y_l^jP_L\tilde{H}^-+g_2P_R\tilde{W}^-].
\end{eqnarray}
The lepton-neutralinos-sleptons are deduced as:
\begin{eqnarray}
&&{\cal L}_{\bar{l}_j\chi^0\tilde{L}}=i\bar{l}_j\Big\{-\Big[\frac{1}{\sqrt{2}}
\Big(2g_1P_L\lambda_{\tilde{B}}+(g_B+2g_{YB})P_L\lambda_{\tilde{B}'}\Big)\tilde{L}^R+Y_l^jP_L\tilde{H}^0_d\tilde{L}^L\Big]\nonumber \\&&\hspace{2.3cm}+\Big[\frac{1}{\sqrt{2}}
\Big(g_2P_R\tilde{W}^0+g_1P_R\lambda_{\tilde{B}}+(g_B+g_{YB})P_R\lambda_{\tilde{B}'}\Big)\tilde{L}^L-Y_l^jP_R\tilde{H}^0_d\tilde{L}^R\Big]\Big\}.
\end{eqnarray}
\subsection{Higgs mass in the B-LSSM}
Due that the strict constraint from SM-like Higgs boson on the numerical results, we discuss the Higgs boson mass matrix. The $\phi_d,\phi_u,\phi_\eta,\phi_{\bar{\eta}}$ mix together at the tree level. In the base $(\phi_d,\phi_u,\phi_\eta,\phi_{\bar{\eta}})$, the tree-level mass squared matrix for neutral CP-even Higgs boson is deduced as:
\begin{eqnarray}
&&\hspace{-0.2cm}M_h^2=\nonumber \\&&\hspace{-0.4cm}u^2\hspace{-0.1cm}\times\hspace{-0.2cm}
\left(\hspace{-0.2cm}\begin{array}{*{20}{c}}
{\frac{1}{4}\frac{g^2 x^2}{1+\tan\beta^2}\hspace{-0.1cm}+\hspace{-0.1cm}n^2\tan\beta}&{-\frac{1}{4}g^2\frac{x^2\tan\beta}{1+\tan\beta^2}}\hspace{-0.1cm}-\hspace{-0.1cm}n^2&
{\frac{1}{2}g_{B}g_{{YB}}\frac{x}{T}}&
{-\frac{1}{2}g_{B}g_{{YB}}\frac{x\times \tan\beta'}{T}}\\ [6pt]
{-\frac{1}{4}g^2\frac{ x^2\tan\beta}{1+\tan\beta^2}}\hspace{-0.1cm}-\hspace{-0.1cm}n^2&{\frac{1}{4}\frac{g^2 x^2\tan\beta^2 }{1+\tan\beta^2}\hspace{-0.1cm}+\hspace{-0.1cm}\frac{n^2}{\tan\beta}}&
{\frac{1}{2}g_{B}g_{{YB}}\frac{x\times \tan\beta}{T}}&{\frac{1}{2}g_{B}g_{{YB}}\frac{x\times \tan\beta\times \tan\beta'}{T}}\\ [6pt]
{\frac{1}{2}g_{B}g_{{YB}}\frac{x}{T}}&{\frac{1}{2}g_{B}g_{{YB}}\frac{x\times \tan\beta}{T}}&{\frac{g_{B}^2}{1+\tan\beta'^2}\hspace{-0.1cm}+\hspace{-0.1cm}\tan\beta'N^2}&
{-g_{B}^2\frac{\tan\beta'}{1+\tan\beta'^2}\hspace{-0.1cm}-\hspace{-0.1cm}N^2}\\ [6pt]
{-\frac{1}{2}g_{B}g_{{YB}}\frac{x\times \tan\beta'}{T}}&{\frac{1}{2}g_{B}g_{{YB}}\frac{x\times \tan\beta\times \tan\beta'}{T}}&
{-g_{B}^2\frac{\tan\beta'}{1+\tan\beta'^2}\hspace{-0.1cm}-\hspace{-0.1cm}N^2}&{g_{B}^2\frac{\tan\beta'^2}{1+\tan\beta'^2}\hspace{-0.1cm}+\hspace{-0.1cm}\frac{N^2}{\tan\beta'}}
\end{array}\hspace{-0.2cm}\right).
\end{eqnarray}
Here, $g^2=g_{1}^2+g_{2}^2+g_{{YB}}^2$, $T=\sqrt{1+\tan\beta^2}\sqrt{1+\tan\beta'^2}$, $x=\frac{v}{u}$, $n^2=\frac{{\rm Re}(B\mu)}{u^2}$ and $N^2=\frac{{\rm Re}(B_\eta)}{u^2}$. The mass of the SM-like Higgs boson can be obtained after considering the leading-log radiative corrections from stop and top particles\cite{leadinglog1,leadinglog2,leadinglog3}.
\begin{eqnarray}
m_h=\sqrt{(m_{h_1}^0)^2+\Delta m_h^2},
\end{eqnarray}
where $m_{h_1}^0$ represents the lightest tree-level Higgs boson mass, and the leading-log radiative corrections $\Delta m_h^2$ can be written as
\begin{eqnarray}
&&\Delta m_h^2=\frac{3m_t^4}{4 \pi ^2 v^2}\Big[\Big(\tilde{t}+\frac{1}{2}\tilde{X}_t\Big)+\frac{1}{16 \pi ^2}\Big(\frac{3m_t^2}{2v^2}-32\pi\alpha_3\Big)(\tilde{t}^2+\tilde{X}_t\tilde{t})\Big],\nonumber \\&&\tilde{t}=\log\frac{M_S^2}{m_t^2},~~~~\tilde{X}_t=\frac{2\tilde{A}_t^2}{M_S^2}(1-\frac{\tilde{A}_t^2}{12M_S^2}).
\end{eqnarray}
Here, $\alpha_3$ is the strong coupling constant, $M_S=\sqrt{m_{\tilde{t}_1}m_{\tilde{t}_2}}$ with $m_{\tilde{t}_{1,2}}$ are the stop masses, $\tilde{A}_t=A_t-\mu_H  \cot \beta $ with $A_t=T_{u,33}$ denotes the trilinear Higgs-stops coupling.
\section{The $(g-2)_{\mu}$ and $l_j^-\rightarrow l_i^- \gamma$ in the B-LSSM}
In this section, we study the muon MDM and the CLFV processes $l_j^-\rightarrow l_i^- \gamma$ in the B-LSSM with the MIA method, which shows the factor of the one-loop contributions more clearly. The concrete contents will be discussed as follows.
\subsection{The one-loop corrections to $(g-2)_{\mu}$ in the B-LSSM}
The effective Lagrangian used here for the muon MDM is given out as follows
\begin{eqnarray}
{\cal L}_{\rm MDM}={e\over4m_{\mu}}\;a_{\mu}\;\bar{l}_\mu\sigma^{\alpha\beta}
l_\mu\;F_{{\alpha\beta}},
\end{eqnarray}
where $\sigma_{\alpha\beta}=i[\gamma_\alpha,\gamma_\beta]/2$, $l_{\mu}$ denotes the wave function of muon, $m_{\mu}$ represents the muon mass, $F_{\alpha\beta}$ is the electromagnetic field strength and $a_\mu$ is the muon MDM.

To obtain the muon MDM, we adopt the effective Lagrangian method\cite{MSSM4,effective1,effective2}, which relates with the following dimension-6 operators.
\begin{eqnarray}
&&\mathcal{O}_1^{\mp}=\frac{1}{(4\pi)^2}\bar{l}(i\mathcal{D}\!\!\!\slash)^3P_{L,R}l,
~~~~~~~~~~~~~~~~\mathcal{O}_2^{\mp}=\frac{eQ_f}{(4\pi)^2}\overline{(i\mathcal{D}_{\mu}l)}\gamma^{\mu}
F\cdot\sigma P_{L,R}l,
\nonumber\\
&&\mathcal{O}_3^{\mp}=\frac{eQ_f}{(4\pi)^2}\bar{l}F\cdot\sigma\gamma^{\mu}
P_{L,R}(i\mathcal{D}_{\mu}l),
~~~~~~\mathcal{O}_4^{\mp}=\frac{eQ_f}{(4\pi)^2}\bar{l}(\partial^{\mu}F_{\mu\nu})\gamma^{\nu}
P_{L,R}l,\nonumber\\&&
\mathcal{O}_5^{\mp}=\frac{m_l}{(4\pi)^2}\bar{l}(i\mathcal{D}\!\!\!\slash)^2P_{L,R}l,
~~~~~~~~~~~~~~~~\mathcal{O}_6^{\mp}=\frac{eQ_fm_l}{(4\pi)^2}\bar{l}F\cdot\sigma
P_{L,R}l,
\end{eqnarray}
with $\mathcal{D}_{\mu}=\partial_{\mu}+ieA_{\mu}$. The operators $\mathcal{O}_{2,3,6}^{\mp}$ have relation with muon MDM when adopting on-shell condition for external leptons. Therefore, we only study the Wilson coefficients $C_{2,3,6}^{\mp}$ of the operators $\mathcal{O}_{2,3,6}^{\mp}$ in the effective Lagrangian. Actually, the Wilson coefficients satisfy the relations $C_{2}^{\mp}=C_{3}^{\mp*}$ and $C_{6}^{+}=C_{6}^{-*}$. Then the muon MDM can be deduced as
\begin{eqnarray}
a_{\mu}=\frac{4Q_fm_{\mu}^2}{(4\pi)^2}\Re(C_2^++C_2^{-*}+C_6^+).
\end{eqnarray}

\begin{figure}[t]
\centering
\includegraphics[width=15cm]{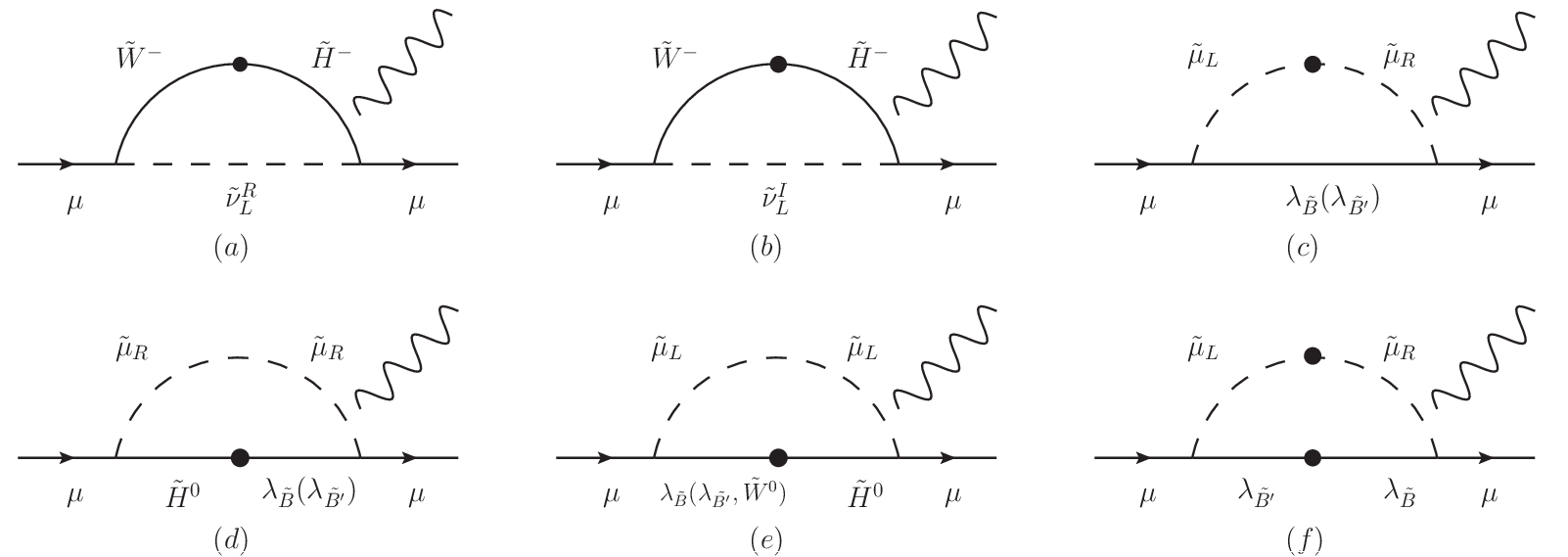}
\caption{The feynman diagrams of muon MDM by MIA in the B-LSSM.} \label{figa}
\end{figure}

The feynman diagrams of muon MDM by MIA in the B-LSSM are shown in FIG.\ref{figa}. Then, we obtain the concrete forms of the one-loop muon MDM in the B-LSSM adopting the MIA method.

1. The one-loop contributions from $\tilde{H}^{-}-\tilde{W}^{-}-\tilde{\nu}_{L}^{R}(\tilde{\nu}_{L}^{I})$:
\begin{eqnarray}
&&a_{\mu}\left(\tilde{\nu}_{L}^{R}, \tilde{H}^{-}, \tilde{W}^{-}\right)= \frac{g_{2}^{2}}{2} x_{\mu} \sqrt{x_{2} x_{\mu_{H}}} \tan \beta\Big[2 \mathcal{I}\left(x_{\mu_{H}}, x_{\tilde{\nu}_{L}^{R}}, x_{2}\right)-\mathcal{J}\left(x_{2}, x_{\mu_{H}}, x_{\tilde{\nu}_{L}^{R}}\right)\nonumber\\
&&~~~~~~~~~~~~~~~~~~~~~~~~~+2 \mathcal{I}\left(x_{2}, x_{\tilde{\nu}_{L}^{R}}, x_{\mu_{H}}\right)-\mathcal{J}\left(x_{\mu_{H}}, x_{2}, x_{\tilde{\nu}_{L}^{R}}\right)\Big], \nonumber\\
&&a_{\mu}\left(\tilde{\nu}_{L}^{I}, \tilde{H}^{-}, \tilde{W}^{-}\right)= \frac{g_{2}^{2}}{2} x_{\mu} \sqrt{x_{2} x_{\mu_{H}}} \tan \beta\Big[2 \mathcal{I}\left(x_{\mu_{H}}, x_{\tilde{\nu}_{L}^{I}}, x_{2}\right)-\mathcal{J}\left(x_{2}, x_{\mu_{H}}, x_{\tilde{\nu}_{L}^{I}}\right)\nonumber\\
&&\hspace{3.5cm}+2 \mathcal{I}\left(x_{2}, x_{\tilde{\nu}_{L}^{I}}, x_{\mu_{H}}\right)-\mathcal{J}\left(x_{\mu_{H}}, x_{2}, x_{\tilde{\nu}_{L}^{I}}\right)\Big],
\label{eq1}
\end{eqnarray}
with $x_{\mu}=\frac{m_{\mu}^2}{\Lambda^{2}}$, $x_{\mu_{H}}=\frac{\mu_H^2}{\Lambda^{2}}$, and $\Lambda$ is the NP energy scale. The one-loop functions $\mathcal{I}(x, y, z)$, $\mathcal{J}(x, y, z)$ and the following $f(x, y, z, t)$, $g(x, y, z, t)$, $k(x, y, z, t)$ are collected in Appendix.

2. The one-loop contributions from $\lambda_{\tilde{B}}\left(\lambda_{\tilde{B'}}\right)-\tilde{\mu}_{L}-\tilde{\mu}_{R}$:
\begin{eqnarray}
&&a_{\mu}\left(\tilde{\mu}_{R}, \tilde{\mu}_{L}, \lambda_{\tilde{B}}\right)=g_{1}^{2} x_{\mu} \sqrt{x_{1} x_{\mu_{H}}} \tan \beta\Big[\mathcal{J}\left(x_{1}, x_{\tilde{\mu}_{L}}, x_{\tilde{\mu}_{R}}\right)+\mathcal{J}\left(x_{1}, x_{\tilde{\mu}_{R}}, x_{\tilde{\mu}_{L}}\right)\Big], \nonumber\\
&&a_{\mu}\left(\tilde{\mu}_{R}, \tilde{\mu}_{L}, \lambda_{\tilde{B'}}\right)=\frac{(g_{B} +2g_{Y B})(g_{B} +g_{Y B})}{2} x_{\mu} \sqrt{x_{B'} x_{\mu_{H}} }\tan \beta\Big[\mathcal{J}\left(x_{B'}, x_{\tilde{\mu}_{L}}, x_{\tilde{\mu}_{R}}\right)\nonumber\\
&&\hspace{3.5cm}+\mathcal{J}\left(x_{B'}, x_{\tilde{\mu}_{R}}, x_{\tilde{\mu}_{L}}\right)\Big] .
\label{eq2}
\end{eqnarray}

3. The one-loop contributions from $\lambda_{\tilde{B}}\left(\lambda_{\tilde{B'}}\right)-\tilde{H}^{0}-\tilde{\mu}_{R}$:
\begin{eqnarray}
&&a_{\mu}\left(\tilde{\mu}_{R}, \lambda_{\tilde{B}}, \tilde{H}^{0}\right)=-g_{1}^{2} x_{\mu} \sqrt{x_{1} x_{\mu_{H}}} \tan \beta\Big[\mathcal{J}\left(x_{1}, x_{\mu_{H}}, x_{\tilde{\mu}_{R}}\right)+\mathcal{J}\left(x_{\mu_{H}}, x_{1}, x_{\tilde{\mu}_{R}}\right)\Big], \nonumber\\
&&a_{\mu}\left(\tilde{\mu}_{R}, \lambda_{\tilde{B'}}, \tilde{H}^{0}\right)=-\frac{(g_{B}+2g_{YB})g_{YB}}{2} x_{\mu} \sqrt{x_{B'} x_{\mu_{H}}} \tan \beta\Big[\mathcal{J}\left(x_{B'}, x_{\mu_{H}}, x_{\tilde{\mu}_{R}}\right)\nonumber\\
&&\hspace{3.5cm}+\mathcal{J}\left(x_{\mu_{H}}, x_{B'}, x_{\tilde{\mu}_{R}}\right)\Big].
\label{eq3}
\end{eqnarray}

4. The one-loop contributions from $\lambda_{\tilde{B}}\left(\tilde{W}^{0}, \lambda_{\tilde{B'}}\right)-\tilde{H}^{0}-\tilde{\mu}_{L}$:
\begin{eqnarray}
&&a_{\mu}\left(\tilde{\mu}_{L}, \tilde{H}^{0},\lambda_{\tilde{B}}\right)= \frac{1}{2} g_{1}^{2} x_{\mu} \sqrt{x_{1} x_{\mu_{H}}} \tan \beta\Big[\mathcal{J}\left(x_{1}, x_{\mu_{H}}, x_{\tilde{\mu}_{L}}\right)+\mathcal{J}\left(x_{\mu_{H}}, x_{1}, x_{\tilde{\mu}_{L}}\right)\Big],\nonumber\\
&&a_{\mu}\left(\tilde{\mu}_{L}, \tilde{H}^{0}, \tilde{W}^{0}\right)=-\frac{1}{2} g_{2}^{2} x_{\mu} \sqrt{x_{2} x_{\mu_{H}}} \tan \beta\Big[\mathcal{J}\left(x_{2}, x_{\mu_{H}}, x_{\tilde{\mu}_{L}}\right)+\mathcal{J}\left(x_{\mu_{H}}, x_{2}, x_{\tilde{\mu}_{L}}\right)\Big],\nonumber\\
&&a_{\mu}\left(\tilde{\mu}_{L}, \tilde{H}^{0}, \lambda_{\tilde{B'}}\right)= \frac{g_{Y B} (g_B+g_{YB})}{2}x_{\mu} \sqrt{x_{B'} x_{\mu_{H}}} \tan \beta\Big[\mathcal{J}\left(x_{B'}, x_{\mu_{H}}, x_{\tilde{\mu}_{L}}\right)\nonumber\\
&&\hspace{3.5cm}+\mathcal{J}\left(x_{\mu_{H}}, x_{B'}, x_{\tilde{\mu}_{L}}\right)\Big] .
\label{eq4}
\end{eqnarray}

5. The one-loop contributions from $\lambda_{\tilde{B}}-\lambda_{\tilde{B'}}-\tilde{\mu}_{R}-\tilde{\mu}_{L}$:
\begin{eqnarray}
&&a_{\mu}\left(\tilde{\mu}_{R}, \tilde{\mu}_{L}, \lambda_{\tilde{B}}, \lambda_{\tilde{B'}}\right)= g_{1}\left(4 g_{Y B}+3g_{B}\right) x_{\mu} \sqrt{x_{B B^{\prime}} x_{\mu_{H}}} \tan \beta \nonumber\\
&&\hspace{3.5cm}\times\Big[\sqrt{x_{1} x_{B'}} f\left(x_{B'}, x_{1}, x_{\tilde{\mu}_{L}}, x_{\tilde{\mu}_{R}}\right)-g\left(x_{B'}, x_{1}, x_{\tilde{\mu}_{L}}, x_{\tilde{\mu}_{R}}\right)\Big] .
\label{eq5}
\end{eqnarray}

In Eqs.(\ref{eq1})-(\ref{eq5}), one can easily find the factors $x_{\mu}$ and $\tan \beta$, which possess the same characteristic as these in MSSM\cite{g-2MIA}. In the B-LSSM, the new gaugino $\lambda_{\tilde{B'}}$ generates the new contributions to muon MDM, which are deduced in  Eqs.(\ref{eq2})-(\ref{eq5}).

To obtain clearer images of the results, we suppose that all the superparticles masses are almost degenerate. The author\cite{g-2MIA} gives the one-loop results of the MSSM in the extreme case where the masses for superparticles $M_{1}, M_{2}, \mu_{H}, m_{\tilde{\mu}_{L}}, m_{\tilde{\mu}_{R}}$ are equal to $\Lambda$
\begin{eqnarray}
a_{\mu}^{\mathrm{MSSM}} \simeq \frac{1}{192 \pi^{2}} \frac{m_{\mu}^{2}}{\Lambda^{2}} \tan \beta\left(5 g_{2}^{2}+g_{1}^{2}\right) .
\end{eqnarray}
Here, we also use the similar case
\begin{eqnarray}
M_{1}=M_{2}=\mu_{H}=m_{\tilde{\nu}_{L}^{R}}=m_{\tilde{\nu}_{L}^{I}}=m_{\tilde{\mu}_{L}}=m_{\tilde{\mu}_{R}}=\left|M_{B'}\right|=\left|M_{B B^{\prime}}\right|=\Lambda.
\end{eqnarray}
Then, the functions $\mathcal{I}(x, y, z), \mathcal{J}(x, y, z), f(x, y, z, t)$ and $ g(x, y, z, t)$ can be simplified as
\begin{eqnarray}
&&\mathcal{I}(1,1,1) =\frac{1}{96 \pi^{2}}, ~~~~~~~~~\mathcal{J}(1,1,1) =\frac{1}{192 \pi^{2}}, \nonumber\\
&&f(1,1,1,1) =-\frac{1}{240 \pi^{2}},~~  g(1,1,1,1) =-\frac{1}{960 \pi^{2}} .
\end{eqnarray}
In this condition, we obtain the simplified one-loop contributions of muon MDM in the B-LSSM.
\begin{eqnarray}
&&a_{\mu}^{B-L}\simeq \frac{1}{960 \pi^{2}} \frac{m_{\mu}^{2}}{\Lambda^{2}} \tan \beta\left[5\left(5 g_{2}^{2}+g_{1}^{2}\right)+5\left(g_{B}^{2}+3g_{YB} g_{B}+g_{YB}^{2}\right) \mathrm{sign}\left[M_{B'}\right]\right.\nonumber\\
&&~~~~~~~~\left.+g_{1}\left(3g_{B}+4g_{YB}\right) \mathrm{sign}\left[M_{B B'}\right]\left(1-4 \mathrm{sign}\left[M_{B'}\right]\right)\right].\label{amuBL}
\end{eqnarray}

In addition to the one-loop MSSM contributions, the Eq.(\ref{amuBL}) adds considerable corrections to $a_{\mu}$ from the new gaugino $\lambda_{\tilde{B'}}$. In the condition $-0.7<g_{Y B}<-0.05$, $0.1<g_{B}<0.85$ and $|g_{Y B}|<g_{B}<\frac{4}{3}|g_{Y B}|$ and with the supposition $\mathrm{sign}\left[M_{B'}\right]=\mathrm{sign}\left[M_{B B'}\right]=-1$, the corrections beyond MSSM can reach large value.
\begin{eqnarray}
a_{\mu}^{B-L} \rightarrow \frac{1}{192 \pi^{2}} \frac{m_{\mu}^{2}}{\Lambda^{2}} \tan \beta\left[(5 g_{2}^{2}+g_{1}^{2})-\left(g_1(3g_B+4g_{YB})+(g_{B}^{2}+3g_{YB} g_{B}+g_{YB}^{2})\right)\right].
\end{eqnarray}
Here, the order analysis shows
\begin{eqnarray}
0<\frac{-g_1(3g_B+4g_{YB})-\left(g_{B}^{2}+3g_{YB} g_{B}+g_{YB}^{2}\right)}{5 g_{2}^{2}+g_{1}^{2}} \lesssim 1.
\end{eqnarray}
Therefore, the B-LSSM contributions beyond MSSM are considerable.
\subsection{The one-loop corrections to $l_j^-\rightarrow l_i^- \gamma$ in the B-LSSM}
\begin{figure}[t]
\centering
\includegraphics[width=14cm]{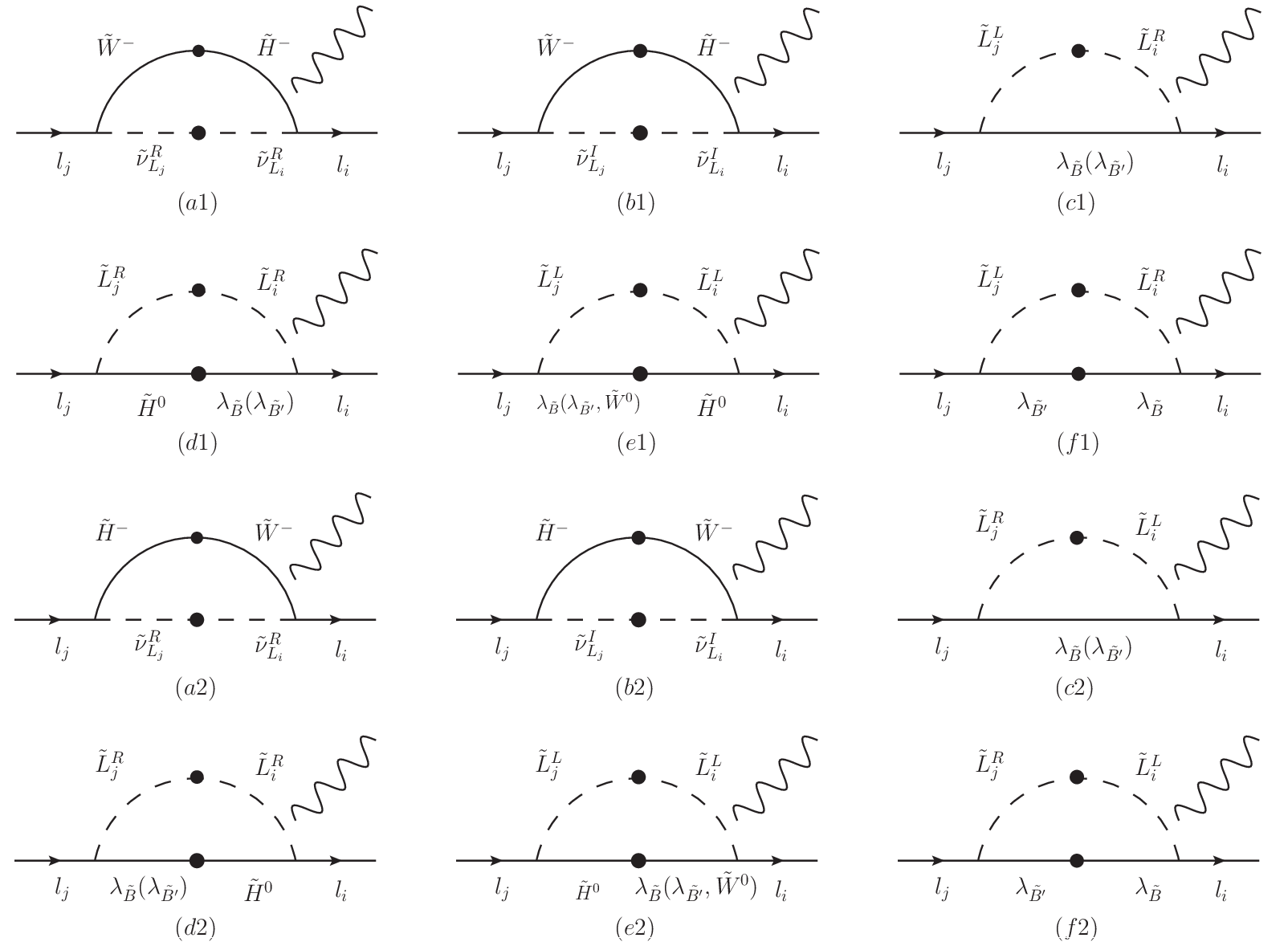}
\caption{The feynman diagrams of $l_j^-\rightarrow l_i^- \gamma$ by MIA in the B-LSSM.} \label{figb}
\end{figure}

Generally, the effective amplitude of the CLFV processes $l_j^-\rightarrow l_i^-\gamma$ can be written as
\begin{eqnarray}
{\cal M}=e\epsilon^{\mu}{\bar u}_i(p+q)[q^2\gamma_{\mu}(C_1^LP_L+C_1^RP_R)+m_{l_j}i\sigma_{\mu\nu}q^{\nu}(C_2^LP_L+C_2^RP_R)]u_j(p).
\end{eqnarray}
where $p$ ($q$) represents the incident lepton (photon) momentum. $m_{l_j}$ is the j-th generation lepton mass. $\epsilon$ is the photon polarization vector and $u_j(p)$ ($\bar{u}_i(p+q)$) is the incident (outgoing) lepton wave function. In FIG.\ref{figb}, we show the relevant B-LSSM Feynman diagrams of $l_j^-\rightarrow l_i^- \gamma$ by MIA. Here, we omit the chirality flips of external lepton in our calculation. In Appendix B, we discuss chirality flips with two examples, and demonstrate that the chirality flips occurring in the internal gaugino lines may yield dominant contributions comparing in the external lepton lines. In Appendix C, we emphasize the contributions from the incident lepton is dominant by comparing the amplitudes of FIG.\ref{figb}(a1) and (a2). The Wilson coefficients $C_{\alpha}^{iL},\; C_{\alpha}^{iR}(\alpha=1,2,\;i=1,\ldots,5)$ adopting the MIA method are obtained from the sum of these diagrams' amplitudes.

1. The one-loop contributions from $\tilde{H}^{-}-\tilde{W}^{-}-\tilde{\nu}_{L}^{R}(\tilde{\nu}_{L}^{I})$:
\begin{eqnarray}
&&C_2^{1L}\left(\tilde{\nu}_{L_j}^{R}, \tilde{\nu}_{L_i}^{R},\tilde{H}^{-}, \tilde{W}^{-}\right)=\frac{g_{2}^{2}}{2\Lambda^4} \frac{m_{l_i}}{m_{l_j}} \sqrt{x_{2} x_{\mu_{H}}} \Delta_{ij}^{LL}\tan \beta\Big[ k(x_{\mu_{H}}, x_{2} , x_{\tilde{\nu}_{L_j}^{R}}, x_{\tilde{\nu}_{L_i}^{R}})\nonumber\\
&&~~~~~~~~~~~~~~~~~~~~~~~~~~~~~~~~+ k(x_{2} ,x_{\mu_{H}}, x_{\tilde{\nu}_{L_j}^{R}}, x_{\tilde{\nu}_{L_i}^{R}})-f(x_{2},x_{\mu_{H}},x_{\tilde{\nu}_{L_j}^{R}}, x_{\tilde{\nu}_{L_i}^{R}})\Big], \nonumber\\
&&C_2^{1L}(\tilde{\nu}_{L_j}^{I}, \tilde{\nu}_{L_i}^{I},\tilde{H}^{-}, \tilde{W}^{-})=\frac{g_{2}^{2}}{2\Lambda^4} \frac{m_{l_i}}{m_{l_j}} \sqrt{x_{2} x_{\mu_{H}}} \Delta_{ij}^{LL}\tan \beta\Big[ k(x_{\mu_{H}}, x_{2} , x_{\tilde{\nu}_{L_j}^{I}}, x_{\tilde{\nu}_{L_i}^{I}})\nonumber\\
&&~~~~~~~~~~~~~~~~~~~~~~~~~~~~~~~+ k(x_{2} ,x_{\mu_{H}}, x_{\tilde{\nu}_{L_j}^{I}}, x_{\tilde{\nu}_{L_i}^{I}})-f(x_{2},x_{\mu_{H}},x_{\tilde{\nu}_{L_j}^{I}}, x_{\tilde{\nu}_{L_i}^{I}})\Big],
\label{eq6}
\end{eqnarray}
\begin{eqnarray}
&&C_2^{1R}\left(\tilde{\nu}_{L_j}^{R}, \tilde{\nu}_{L_i}^{R},\tilde{H}^{-}, \tilde{W}^{-}\right)=\frac{g_{2}^{2}}{2\Lambda^4} \sqrt{x_{2} x_{\mu_{H}}} \Delta_{ij}^{LL}\tan \beta\Big[ k(x_{\mu_{H}}, x_{2} , x_{\tilde{\nu}_{L_j}^{R}}, x_{\tilde{\nu}_{L_i}^{R}})\nonumber\\
&&~~~~~~~~~~~~~~~~~~~~~~~~~~~~~~~~+ k(x_{2} ,x_{\mu_{H}}, x_{\tilde{\nu}_{L_j}^{R}}, x_{\tilde{\nu}_{L_i}^{R}})-f(x_{2},x_{\mu_{H}},x_{\tilde{\nu}_{L_j}^{R}}, x_{\tilde{\nu}_{L_i}^{R}})\Big], \nonumber\\
&&C_2^{1R}(\tilde{\nu}_{L_j}^{I}, \tilde{\nu}_{L_i}^{I},\tilde{H}^{-}, \tilde{W}^{-})=\frac{g_{2}^{2}}{2\Lambda^4} \sqrt{x_{2} x_{\mu_{H}}} \Delta_{ij}^{LL}\tan \beta\Big[ k(x_{\mu_{H}}, x_{2} , x_{\tilde{\nu}_{L_j}^{I}}, x_{\tilde{\nu}_{L_i}^{I}})\nonumber\\
&&~~~~~~~~~~~~~~~~~~~~~~~~~~~~~~~+ k(x_{2} ,x_{\mu_{H}}, x_{\tilde{\nu}_{L_j}^{I}}, x_{\tilde{\nu}_{L_i}^{I}})-f(x_{2},x_{\mu_{H}},x_{\tilde{\nu}_{L_j}^{I}}, x_{\tilde{\nu}_{L_i}^{I}})\Big].
\label{eq61}
\end{eqnarray}

2. The one-loop contributions from $\lambda_{\tilde{B}}\left(\lambda_{\tilde{B'}}\right)-\tilde{L}_j^L-\tilde{L}_i^{R}$:
\begin{eqnarray}
&&C_2^{2L}\left(\tilde{L}_j^L,\tilde{L}_i^{R}, \lambda_{\tilde{B}}\right)=-\frac{g_{1}^{2}} {2\Lambda^3 m_{l_j}} \sqrt{x_{1}} \Delta_{ij}^{LR}\Big[\mathcal{J}(x_{1}, x_{\tilde{L}_j^L},x_{\tilde{L}_i^{R}})+\mathcal{J}(x_{1}, x_{\tilde{L}_i^{R}}, x_{\tilde{L}_j^L})\Big] , \nonumber\\
&&C_2^{2L}\left(\tilde{L}_j^L,\tilde{L}_i^{R}, \lambda_{\tilde{B'}}\right)=-\frac{(g_{B} +2g_{Y B})(g_{B} +g_{Y B})}{4\Lambda^3 m_{l_j}}  \sqrt{x_{B'} } \Delta_{ij}^{LR}\Big[\mathcal{J}(x_{B'}, x_{\tilde{L}_j^L},x_{\tilde{L}_i^{R}})\nonumber\\
&&\hspace{3.5cm}+\mathcal{J}(x_{B'}, x_{\tilde{L}_i^{R}}, x_{\tilde{L}_j^L})\Big],
\label{eq7}
\end{eqnarray}
\begin{eqnarray}
&&C_2^{2R}\left(\tilde{L}_j^R,\tilde{L}_i^{L}, \lambda_{\tilde{B}}\right)=-\frac{g_{1}^{2}} {2\Lambda^3 m_{l_j}} \sqrt{x_{1}} \Delta_{ij}^{LR}\Big[\mathcal{J}(x_{1}, x_{\tilde{L}_j^R},x_{\tilde{L}_i^{ L}})+\mathcal{J}(x_{1}, x_{\tilde{L}_i^{ L}}, x_{\tilde{L}_j^R})\Big] , \nonumber\\
&&C_2^{2R}\left(\tilde{L}_j^R,\tilde{L}_i^{L}, \lambda_{\tilde{B'}}\right)=-\frac{(g_{B} +2g_{Y B})(g_{B} +g_{Y B})}{4\Lambda^3 m_{l_j}}  \sqrt{x_{B'} } \Delta_{ij}^{LR}\Big[\mathcal{J}(x_{B'}, x_{\tilde{L}_j^R},x_{\tilde{L}_i^{ L}})\nonumber\\
&&\hspace{3.5cm}+\mathcal{J}(x_{B'}, x_{\tilde{L}_i^{L}}, x_{\tilde{L}_j^R})\Big] .
\label{eq71}
\end{eqnarray}

3. The one-loop contributions from $\lambda_{\tilde{B}}\left(\lambda_{\tilde{B'}}\right)-\tilde{H}^{0}-\tilde{L}_j^{R}-\tilde{L}_i^{R}$:
\begin{eqnarray}
&&C_2^{3L}\left(\tilde{L}_j^{R},\tilde{L}_i^{R}, \lambda_{\tilde{B}}, \tilde{H}^{0}\right)=-\frac{g_{1}^{2}}{\Lambda^4} \Delta_{ij}^{RR}\hspace{-0.1cm}\Big[\hspace{-0.1cm}\sqrt{x_{1} x_{\mu_{H}}} \tan \beta f(x_{1}, x_{\mu_{H}}, x_{\tilde{L}_j^{R}}, x_{\tilde{L}_i^{R}})\hspace{-0.1cm}-\hspace{-0.1cm}g(x_{1}, x_{\mu_{H}}, x_{\tilde{L}_j^{R}}, x_{\tilde{L}_i^{R}})\Big], \nonumber\\
&&C_2^{3L}\left(\tilde{L}_j^{R},\tilde{L}_i^{R}, \lambda_{\tilde{B'}}, \tilde{H}^{0}\right)=-\frac{(g_{B}+2g_{YB})g_{YB}}{2\Lambda^4} \Delta_{ij}^{RR}\Big[\sqrt{x_{B'} x_{\mu_{H}}} \tan \beta f(x_{B'}, x_{\mu_{H}}, x_{\tilde{L}_j^{R}}, x_{\tilde{L}_i^{R}})\nonumber\\
&&\;\;\;\;\;\;\;\;\;\;\;\;\;\;\;\;\;\;\;\;\;\;\;\;\;\;\;\;\;\;\;\;\;\;\;\;\;-g(x_{B'}, x_{\mu_{H}}, x_{\tilde{L}_j^{R}}, x_{\tilde{L}_i^{R}})\Big],
\label{eq8}
\end{eqnarray}
\begin{eqnarray}
&&C_2^{3R}\left(\tilde{L}_j^{R},\tilde{L}_i^{R}, \lambda_{\tilde{B}},\tilde{H}^{0}\right)=-\frac{g_{1}^{2}}{\Lambda^4} \frac{m_{l_i}}{m_{l_j}}\Delta_{ij}^{RR}\hspace{-0.1cm}\Big[\hspace{-0.1cm}\sqrt{x_{1} x_{\mu_{H}}} \tan \beta f(x_{1}, x_{\mu_{H}}, x_{\tilde{L}_j^{R}}, x_{\tilde{L}_i^{R}})\nonumber\\
&&\;\;\;\;\;\;\;\;\;\;\;\;\;\;\;\;\;\;\;\;\;\;\;\;\;\;\;\;\;\;\;\;\;\;\;\;\;-g(x_{1}, x_{\mu_{H}}, x_{\tilde{L}_j^{R}}, x_{\tilde{L}_i^{R}})\Big], \nonumber\\
&&C_2^{3R}\left(\tilde{L}_j^{R},\tilde{L}_i^{R},\lambda_{\tilde{B'}}, \tilde{H}^{0} \right)=-\frac{(g_{B}+2g_{YB})g_{YB}}{2\Lambda^4}\frac{m_{l_i}}{m_{l_j}} \Delta_{ij}^{RR}\Big[\sqrt{x_{B'} x_{\mu_{H}}} \tan \beta f(x_{B'}, x_{\mu_{H}}, x_{\tilde{L}_j^{R}}, x_{\tilde{L}_i^{R}})\nonumber\\
&&\;\;\;\;\;\;\;\;\;\;\;\;\;\;\;\;\;\;\;\;\;\;\;\;\;\;\;\;\;\;\;\;\;\;\;\;\;-g(x_{B'}, x_{\mu_{H}}, x_{\tilde{L}_j^{R}}, x_{\tilde{L}_i^{R}})\Big].
\label{eq81}
\end{eqnarray}

4. The one-loop contributions from $\lambda_{\tilde{B}}\left(\tilde{W}^{0}, \lambda_{\tilde{B'}}\right)-\tilde{H}^{0}-\tilde{L}_j^{L}-\tilde{L}_i^{L}$:
\begin{eqnarray}
&&C_2^{4L}\left(\tilde{L}_j^{L},\tilde{L}_i^{L}, \tilde{H}^{0},\lambda_{\tilde{B}}\right)= \frac{g_{1}^{2}}{2\Lambda^4}\frac{m_{l_i}}{m_{l_j}} \Delta_{ij}^{LL}\Big[\sqrt{x_{1} x_{\mu_{H}}} \tan \beta f(x_{1}, x_{\mu_{H}}, x_{\tilde{L}_j^{L}}, x_{\tilde{L}_i^{L}})\nonumber\\
&&\;\;\;\;\;\;\;\;\;\;\;\;\;\;\;\;\;\;\;\;\;\;\;\;\;\;\;\;\;\;\;\;\;\;\;\;-g(x_{1}, x_{\mu_{H}}, x_{\tilde{L}_j^{L}}, x_{\tilde{L}_i^{L}})\Big],\nonumber\\
&&C_2^{4L}\left(\tilde{L}_j^{L},\tilde{L}_i^{L}, \tilde{H}^{0}, \tilde{W}^{0}\right)=-\frac{ g_{2}^{2}}{2\Lambda^4}\frac{m_{l_i}}{m_{l_j}} \Delta_{ij}^{LL} \Big[\sqrt{x_{2} x_{\mu_{H}}} \tan \beta f(x_{2}, x_{\mu_{H}}, x_{\tilde{L}_j^{L}}, x_{\tilde{L}_i^{L}})\nonumber\\
&&\;\;\;\;\;\;\;\;\;\;\;\;\;\;\;\;\;\;\;\;\;\;\;\;\;\;\;\;\;\;\;\;\;\;\;\;\;-g(x_{2}, x_{\mu_{H}}, x_{\tilde{L}_j^{L}}, x_{\tilde{L}_i^{L}})\Big],\nonumber\\
&&C_2^{4L}\left(\tilde{L}_j^{L},\tilde{L}_i^{L}, \tilde{H}^{0}, \lambda_{\tilde{B'}}\right)= \frac{g_{Y B} (g_B+g_{YB})}{2\Lambda^4}\frac{m_{l_i}}{m_{l_j}} \Delta_{ij}^{LL} \Big[\sqrt{x_{B'} x_{\mu_{H}}} \tan \beta f(x_{B'}, x_{\mu_{H}}, x_{\tilde{L}_j^{L}}, x_{\tilde{L}_i^{L}})\nonumber\\
&&\;\;\;\;\;\;\;\;\;\;\;\;\;\;\;\;\;\;\;\;\;\;\;\;\;\;\;\;\;\;\;\;\;\;\;\;\;-g(x_{B'}, x_{\mu_{H}}, x_{\tilde{L}_j^{L}}, x_{\tilde{L}_i^{L}})\Big],
\label{eq9}
\end{eqnarray}
\begin{eqnarray}
&&C_2^{4R}\left(\tilde{L}_j^{L},\tilde{L}_i^{L}, \tilde{H}^{0},\lambda_{\tilde{B}}\right)= \frac{g_{1}^{2}}{2\Lambda^4} \Delta_{ij}^{LL}\Big[\sqrt{x_{1} x_{\mu_{H}}} \tan \beta f(x_{1}, x_{\mu_{H}}, x_{\tilde{L}_j^{L}}, x_{\tilde{L}_i^{L}})\nonumber\\
&&\;\;\;\;\;\;\;\;\;\;\;\;\;\;\;\;\;\;\;\;\;\;\;\;\;\;\;\;\;\;\;\;\;\;\;\;-g(x_{1}, x_{\mu_{H}}, x_{\tilde{L}_j^{L}}, x_{\tilde{L}_i^{L}})\Big],\nonumber\\
&&C_2^{4R}\left(\tilde{L}_j^{L},\tilde{L}_i^{L}, \tilde{H}^{0}, \tilde{W}^{0}\right)=-\frac{ g_{2}^{2}}{2\Lambda^4}\Delta_{ij}^{LL} \Big[\sqrt{x_{2} x_{\mu_{H}}} \tan \beta f(x_{2}, x_{\mu_{H}}, x_{\tilde{L}_j^{L}}, x_{\tilde{L}_i^{L}})\nonumber\\
&&\;\;\;\;\;\;\;\;\;\;\;\;\;\;\;\;\;\;\;\;\;\;\;\;\;\;\;\;\;\;\;\;\;\;\;\;\;-g(x_{2}, x_{\mu_{H}}, x_{\tilde{L}_j^{L}}, x_{\tilde{L}_i^{L}})\Big],\nonumber\\
&&C_2^{4R}\left(\tilde{L}_j^{L},\tilde{L}_i^{L}, \tilde{H}^{0}, \lambda_{\tilde{B'}}\right)= \frac{g_{Y B} (g_B+g_{YB})}{2\Lambda^4}\Delta_{ij}^{LL} \Big[\sqrt{x_{B'} x_{\mu_{H}}} \tan \beta f(x_{B'}, x_{\mu_{H}}, x_{\tilde{L}_j^{L}}, x_{\tilde{L}_i^{L}})\nonumber\\
&&\;\;\;\;\;\;\;\;\;\;\;\;\;\;\;\;\;\;\;\;\;\;\;\;\;\;\;\;\;\;\;\;\;\;\;\;\;-g(x_{B'}, x_{\mu_{H}}, x_{\tilde{L}_j^{L}}, x_{\tilde{L}_i^{L}})\Big] .
\label{eq91}
\end{eqnarray}

5. The one-loop contributions from $\lambda_{\tilde{B}}-\lambda_{\tilde{B'}}-\tilde{L}_j^{L}-\tilde{L}_i^{R}$:
\begin{eqnarray}
&&C_2^{5L}\left(\tilde{L}_j^{L},\tilde{L}_i^{R}, \lambda_{\tilde{B}}, \lambda_{\tilde{B'}}\right)=-\frac{g_{1}(4 g_{Y B}+3g_{B})}{2\Lambda^3m_{l_j}} \sqrt{x_{B B^{\prime}}}\Delta_{ij}^{LR}\Big[\sqrt{x_{1} x_{B'}} f(x_{B'},x_{1}, x_{\tilde{L}_j^{L}}, x_{\tilde{L}_i^{R}})\nonumber\\
&&\;\;\;\;\;\;\;\;\;\;\;\;\;\;\;\;\;\;\;\;\;\;\;\;\;\;\;\;\;\;\;\;\;\;\;\;\;-g(x_{B'},x_{1}, x_{\tilde{L}_j^{L}}, x_{\tilde{L}_i^{R}})\Big],
\label{eq10}
\end{eqnarray}
\begin{eqnarray}
&&C_2^{5R}\left(\tilde{L}_j^{R},\tilde{L}_i^{L}, \lambda_{\tilde{B}}, \lambda_{\tilde{B'}}\right)=-\frac{g_{1}(4 g_{Y B}+3g_{B})}{2\Lambda^3m_{l_j}} \sqrt{x_{B B^{\prime}}}\Delta_{ij}^{LR}\Big[\sqrt{x_{1} x_{B'}} f(x_{B'},x_{1}, x_{\tilde{L}_j^{R}}, x_{\tilde{L}_i^{ L}})\nonumber\\
&&\;\;\;\;\;\;\;\;\;\;\;\;\;\;\;\;\;\;\;\;\;\;\;\;\;\;\;\;\;\;\;\;\;\;\;\;\;-g(x_{B'},x_{1}, x_{\tilde{L}_j^{R}}, x_{\tilde{L}_i^{L }})\Big] .
\label{eq101}
\end{eqnarray}

Then the decay width of $l_j^-\rightarrow l_i^- \gamma$ can be written as
\begin{eqnarray}
\Gamma(l_j^-\rightarrow l_i^- \gamma)=\frac{e^2}{16\pi}m_{l_j}^5(|C_{2}^L|^2+|C_{2}^R|^2),
\end{eqnarray}
with the final Wilson coefficient $C_{2}^{L,R}=\sum_i^{i=1\cdots5} C_{2}^{iL,iR},\;i=1,\cdot\cdot\cdot,5$. And the branching ratio for $l_j^-\rightarrow l_i^- \gamma$ is
\begin{eqnarray}
Br(l_j^-\rightarrow l_i^- \gamma)=\Gamma(l_j^-\rightarrow l_i^- \gamma)/\Gamma_{l_j}.
\end{eqnarray}

From the Eqs.(\ref{eq6})-(\ref{eq101}), we find that $C_2^{L,R}$ are almost affected by $\tan\beta$ and $\Delta_{ij}^{AB}(A,B=L,R)$. Here, $\Delta_{ij}^{LL}=m_{{L}}^2\delta_{ij}^{LL},\Delta_{ij}^{RR}=m_{{E}}^2\delta_{ij}^{RR}$ and $
\Delta_{ij}^{LR}=A_e v_d \delta_{ij}^{LR}$, which are related with soft breaking slepton mass squared matrices $m_{\tilde{L},\tilde{E}}^2$ and the trilinear coupling matrix $T_e$, whose off-diagonal terms introduce the slepton flavor mixing.
\begin{eqnarray}
m_{\tilde{L}}^2\hspace{-0.1cm}=\hspace{-0.2cm}\left(\hspace{-0.1cm}\begin{array}{ccc}
1, & \delta_{12}^{LL}, &  \delta_{13}^{LL}\\
 \delta_{12}^{LL}, & 1, & \delta_{23}^{LL}\\
 \delta_{13}^{LL}, & \delta_{23}^{LL}, &1
\end{array}\hspace{-0.1cm}\right)\hspace{-0.1cm}m_{L}^2,~~~
m_{\tilde{E}}^2\hspace{-0.1cm}=\hspace{-0.2cm}\left(\hspace{-0.1cm}\begin{array}{ccc}
1, & \delta_{12}^{RR}, &  \delta_{13}^{RR}\\
 \delta_{12}^{RR}, & 1, & \delta_{23}^{RR}\\
 \delta_{13}^{RR}, & \delta_{23}^{RR}, &1
\end{array}\hspace{-0.1cm}\right)\hspace{-0.1cm}m_{E}^2,~~~
T_e\hspace{-0.1cm}=\hspace{-0.2cm}\left(\hspace{-0.1cm}\begin{array}{ccc}
1, & \delta_{12}^{LR}, &  \delta_{13}^{LR}\\
 \delta_{12}^{LR}, & 1, & \delta_{23}^{LR}\\
 \delta_{13}^{LR}, & \delta_{23}^{LR}, &1
\end{array}\hspace{-0.1cm}\right)\hspace{-0.1cm}A_e,
\end{eqnarray}
In the subsequent numerical analyses, we discuss the branching ratios for CLFV processes $l_j^-\rightarrow l_i^- \gamma$ in the B-LSSM depending on the slepton mixing parameters.

In order to more intuitively analyze the factors that affect CLFV processes $l_j^-\rightarrow l_i^- \gamma$, we also suppose that the superparticles masses are almost degenerate. In other words, we give the one-loop results in the extreme case where the super particle masses are equal to $\Lambda$.
\begin{eqnarray}
|M_{1}|=|M_{2}|=|\mu_{H}|=m_{{L}}=m_{{E}}=\left|M_{B'}\right|=\left|M_{B B^{\prime}}\right|=\Lambda.
\end{eqnarray}
Here, the function $k(x, y, z, t)$ is much simplified as $k(1,1,1,1) =-\frac{1}{192 \pi^{2}}$, then the Wilson coefficients from the MSSM can be deduced as
\begin{eqnarray}
&&C_2^{\mathrm{L,MSSM}} \simeq \frac{1}{1920 \pi^{2}\Lambda^4 m_{l_j}}\Big[-10 \sqrt{2} g_1^2 \Lambda A_e v_d \mathrm{sign}\left[M_{1}\right]\delta_{ij}^{LR}\nonumber\\
&&\;\;\;\;\;\;\;\;\;\;\;\;\;\;\;\;+2 g_1^2(4\tan\beta\mathrm{sign}\left[M_{1} \mu_H\right]-1) m_{l_j}m_{E}^2 \delta_{ij}^{RR}\Big],
\end{eqnarray}
\begin{eqnarray}
&&C_2^{\mathrm{R,MSSM}} \simeq \frac{1}{1920 \pi^{2}\Lambda^4 m_{l_j}}\Big[-10 \sqrt{2} g_1^2 \Lambda A_e v_d \mathrm{sign}\left[M_{1}\right]\delta_{ij}^{LR}- \Big(g_2^2(8\tan\beta\mathrm{sign}\left[M_{2} \mu_H\right]+1)\nonumber\\
&&\;\;\;\;\;\;\;\;\;\;\;\;\;\;\;\;+g_1^2(4\tan\beta\mathrm{sign}\left[M_{1} \mu_H\right]-1)\Big) m_{l_j} m_{{L}}^2 \delta_{ij}^{LL}\Big].
\end{eqnarray}
With the supposition $\mathrm{sign}\left[M_{1}\right]=\mathrm{sign}\left[\mu_H\right]=\mathrm{sign}\left[M_{2}\right]=1$ and $4\tan\beta>>1$, the one-loop corrections to the Wilson coefficient of MSSM can reach
\begin{eqnarray}
&&C_2^{\mathrm{L,MSSM}} \rightarrow\frac{1}{1920 \pi^{2}\Lambda^2 m_{l_j}}\Big[-10 \sqrt{2} g_1^2 \frac{A_e v_d}{\Lambda}\delta_{ij}^{LR}+8 g_1^2\tan\beta m_{l_j} \delta_{ij}^{RR}\Big],
\end{eqnarray}
\begin{eqnarray}
&&C_2^{\mathrm{R,MSSM}} \rightarrow\frac{1}{1920 \pi^{2}\Lambda^2 m_{l_j}}\Big[-10 \sqrt{2} g_1^2 \frac{A_e v_d}{\Lambda}\delta_{ij}^{LR}- \big(8g_2^2+4g_1^2\big)\tan\beta m_{l_j}\delta_{ij}^{LL}\Big] .
\end{eqnarray}

In the same way, the Wilson coefficient from the B-LSSM can be written as
\begin{eqnarray}
&&C_2^{\mathrm{L,B-L}} \simeq -\frac{1}{3840 \pi^{2}\Lambda^4 m_{l_j}}\Big[\sqrt{2} \Big(20 g_1^2\mathrm{sign}\left[M_{1}\right]+10(g_B+g_{YB})(g_B+2g_{YB})\mathrm{sign}\left[M_{B'}\right]\nonumber\\
&&-g_1(3g_B+4g_{YB})\mathrm{sign}\left[M_{BB'}\right] (4 \mathrm{sign}\left[M_{1}M_{B'}\right]-1) \Big)\Lambda A_e v_d \delta_{ij}^{LR}-2 \Big(g_1^2(8\tan\beta\mathrm{sign}\left[\mu_H M_{1} \right]\nonumber\\
&&-2) +g_{YB}(g_B+2g_{YB})(4\tan\beta \mathrm{sign}\left[\mu_H M_{B'}\right]-1)\Big)m_{l_j} m_{{E}}^2 \delta_{ij}^{RR}\Big],
\end{eqnarray}
\begin{eqnarray}
&&C_2^{\mathrm{R,B-L}} \simeq -\frac{1}{3840 \pi^{2}\Lambda^4 m_{l_j}}\Big[\sqrt{2} \Big(20 g_1^2\mathrm{sign}\left[M_{1}\right]+10(g_B+g_{YB})(g_B+2g_{YB})\mathrm{sign}\left[M_{B'}\right]\nonumber\\
&&\;\;\;\;\;\;\;\;\;-g_1(3g_B+4g_{YB})\mathrm{sign}\left[M_{BB'}\right] (4 \mathrm{sign}\left[M_{1}M_{B'}\right]-1) \Big)\Lambda A_e v_d \delta_{ij}^{LR}\nonumber\\
&&\;\;\;\;\;\;\;\;\;+ 2\Big(g_2^2(8\tan\beta\mathrm{sign}\left[\mu_H M_{2} \right]+1)+g_1^2(4\tan\beta\mathrm{sign}\left[\mu_H M_{1} \right]-1) \nonumber\\
&&\;\;\;\;\;\;\;\;\;+g_{YB}(g_B+g_{YB})(4\tan\beta \mathrm{sign}\left[\mu_H M_{B'}\right]-1)\Big) m_{l_j} m_{{L}}^2 \delta_{ij}^{LL}\Big].
\end{eqnarray}
With the supposition $\mathrm{sign}\left[M_{1}\right]=\mathrm{sign}\left[M_{2}\right]=\mathrm{sign}\left[\mu_H\right]=1$, $\mathrm{sign}\left[M_{B'}\right]=\mathrm{sign}\left[M_{BB'}\right]=-1$ and $4\tan\beta>>1$, the Wilson coefficient for the B-LSSM can be
\begin{eqnarray}
&&C_2^{\mathrm{L,B-L}} \rightarrow \frac{1}{3840 \pi^{2}\Lambda^2 m_{l_j}}\Big[\hspace{-0.1cm}-\hspace{-0.1cm}20\sqrt{2} g_1^2 \frac{A_e v_d}{\Lambda}\delta_{ij}^{LR}+16 g_1^2\tan\beta m_{l_j} \delta_{ij}^{RR}\nonumber\\
&&\;\;\;\;\;\;\;\;\;\;\;\;\;\;\;\;+\sqrt{2} \Big(10(g_B+g_{YB})(g_B+2g_{YB})+5g_1(3g_B+4g_{YB})\Big) \frac{A_e v_d}{\Lambda}\delta_{ij}^{LR}\nonumber\\
&&\;\;\;\;\;\;\;\;\;\;\;\;\;\;\;\;-8 g_{YB}(g_B+2g_{YB})\tan\beta m_{l_j} \delta_{ij}^{RR}\Big],
\end{eqnarray}
\begin{eqnarray}
&&C_2^{\mathrm{R,B-L}} \rightarrow \frac{1}{3840 \pi^{2}\Lambda^2 m_{l_j}}\Big[\hspace{-0.1cm}-\hspace{-0.1cm}20\sqrt{2} g_1^2 \frac{A_e v_d}{\Lambda}\delta_{ij}^{LR}- 2\big(8g_2^2+4g_1^2\big)\tan\beta m_{l_j} \delta_{ij}^{LL}\nonumber\\
&&\;\;\;\;\;\;\;\;\;\;\;\;\;\;\;\;+\sqrt{2} \Big(10(g_B+g_{YB})(g_B+2g_{YB})+5g_1(3g_B+4g_{YB})\Big) \frac{A_e v_d}{\Lambda}\delta_{ij}^{LR}\nonumber\\
&&\;\;\;\;\;\;\;\;\;\;\;\;\;\;\;\;+ 8g_{YB}(g_B+g_{YB})\tan\beta m_{l_j} \delta_{ij}^{LL}\Big] .
\end{eqnarray}
In the condition $-0.7<g_{Y B}<-0.05$, $0.1<g_{B}<0.85$ and $|g_{Y B}|<g_{B}<4/3|g_{Y B}|$, then $\sqrt{2} \Big(10(g_B+g_{YB})(g_B+2g_{YB})+5g_1(3g_B+4g_{YB})\Big)<0$, $-8 g_{YB}(g_B+2g_{YB})<0$, $8g_{YB}(g_B+g_{YB})<0$. The order analysis shows
\begin{eqnarray}
&&D_2^{L,B-L}=\Big[\sqrt{2} \Big(10(g_B+g_{YB})(g_B+2g_{YB})+5g_1(3g_B+4g_{YB})\Big) \frac{A_e v_d}{\Lambda}\delta_{ij}^{LR}\nonumber\\
&&-8g_{YB}(g_B+2g_{YB})\tan\beta m_{l_j} \delta_{ij}^{RR}\Big]<0,
\end{eqnarray}
\begin{eqnarray}
&&D_2^{R,B-L}=\Big[\sqrt{2} \Big(10(g_B+g_{YB})(g_B+2g_{YB})+5g_1(3g_B+4g_{YB})\Big) \frac{A_e v_d}{\Lambda}\delta_{ij}^{LR}\nonumber\\
&&+8g_{YB}(g_B+g_{YB})\tan\beta m_{l_j} \delta_{ij}^{LL}\Big]<0.
\end{eqnarray}
As parameters $g_1\simeq0.3$, $g_2\simeq0.6$, $g_B\simeq0.3$, $g_{YB}\simeq-0.25$, $\tan\beta m_{\mu}\simeq\frac{A_e v_d}{\Lambda}\simeq1$ and $\delta_{12}^{LL}\simeq\delta_{12}^{RR}\simeq\delta_{12}^{LR}\simeq10^{-5}$, $D_2^{L,B-L}$ and $ D_2^{R,B-L}$ are at the order around $-10^{-5}$, which approximately takes the same values as the MSSM contributions from $[-10 \sqrt{2} g_1^2 \frac{A_e v_d}{\Lambda}\delta_{ij}^{LR}+8 g_1^2\tan\beta m_{l_j} \delta_{ij}^{RR}]$ and $[-10 \sqrt{2} g_1^2 \frac{A_e v_d}{\Lambda}\delta_{ij}^{LR}- \big(8g_2^2+4g_1^2\big)\tan\beta m_{l_j}\delta_{ij}^{LL}]$. Therefore, the B-LSSM contributions beyond MSSM are considerable.
\section{The numerical results}
In this section, we research the numerical results of the branching ratios for CLFV processes $l_j^-\rightarrow l_i^- \gamma$. We consider some experimental constraints:
1. The updated experimental data indicates that the $Z'$ boson mass satisfies $M'_Z\geq 5.15~ {\rm TeV}$ with $95\%$ CL\cite{Zpupper}. Refs.\cite{Zpupper1,Zpupper2} give us an upper bound on the ratio between the mass of $Z'$ boson and its gauge coupling at $99\%$ CL: $\frac{M'_Z}{g_B}\geq 6~{\rm TeV}$, so $g_B$ is restricted in the region of $0 < g_B \leq0.85$. 2. The large $\tan\beta$ has been excluded by the $\bar{B}\rightarrow X_s\gamma$ experiment\cite{BSgamma1,BSgamma2}. Besides, the Yukawa coupling $Y_b$ is defined as $Y_b=\sqrt{2(\tan\beta^2+1)}m_b/v$. In general, the value of $Y_b$ is smaller than 1, $m_b\simeq4.18~\rm GeV$ and $v\simeq246~\rm GeV$, so the parameter $\tan\beta$ should be approximatively smaller than 40. 3. The coupling parameter $g_{YB}$ is taken around $-0.45\sim-0.05$, which has been discussed in the works\cite{gYB1,gYB2}.

The SM-like Higgs boson mass is $m_h=125.25\pm0.17$ GeV\cite{PDG2022,Higgsmassexp1,Higgsmassexp2,Higgsmassexp3}, which constrains the parameter space strictly. Therefore, we take the suitable parameter space to limit the SM-like Higgs boson mass of the B-LSSM within experimental $4\sigma,3\sigma$ and $2\sigma$ regions. We also consider the constraints on the NP contribution to the muon MDM in the B-LSSM. The difference between the experimental measurement and SM theoretical prediction of $a_{\mu}$ is $\Delta a_{\mu} =a^{exp}_{\mu}-a^{SM}_{\mu}=(251\pm59)\times 10^{-11}$\cite{PDG2022,g-2exp}, which possesses $4.2\sigma$ deviation. Therefore, we constrain the NP contribution $\Delta a_{\mu}^{NP}$ with $4\sigma,3\sigma$ and $2\sigma$ experimental errors. Then the numerical results of the CLFV process $l_j^-\rightarrow l_i^- \gamma$ are discussed detailedly in the follows.
\begin{figure}[t]
\centering
\includegraphics[width=7cm]{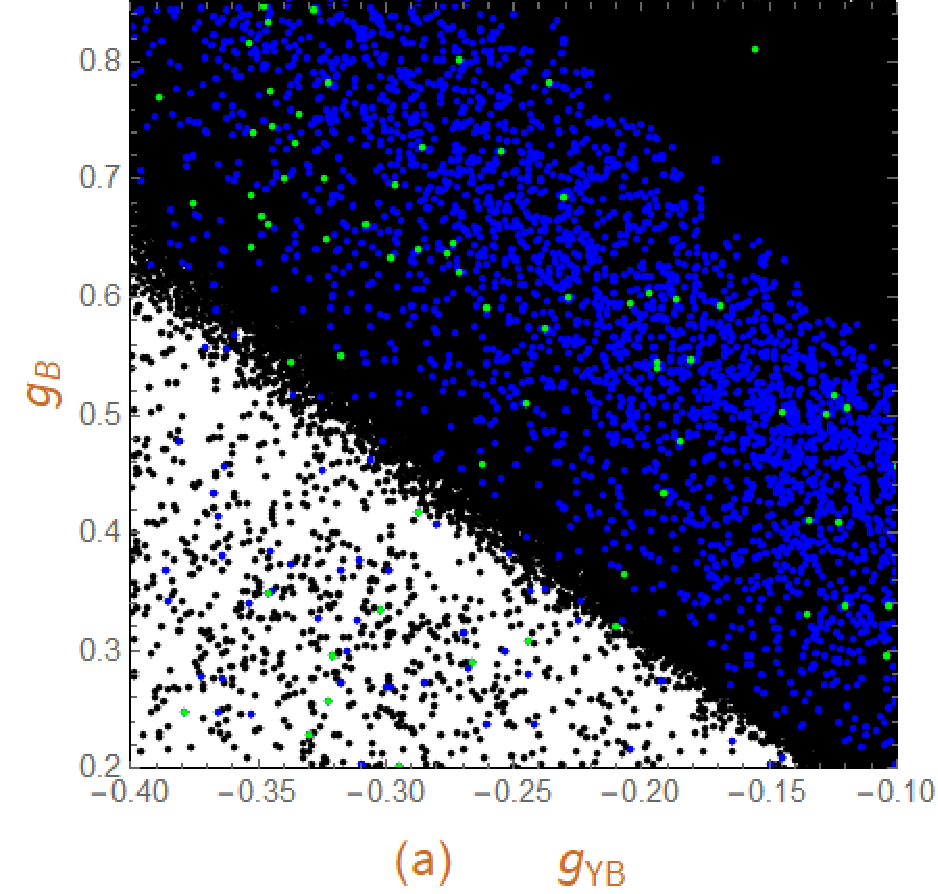}~~~~
\includegraphics[width=7.1cm]{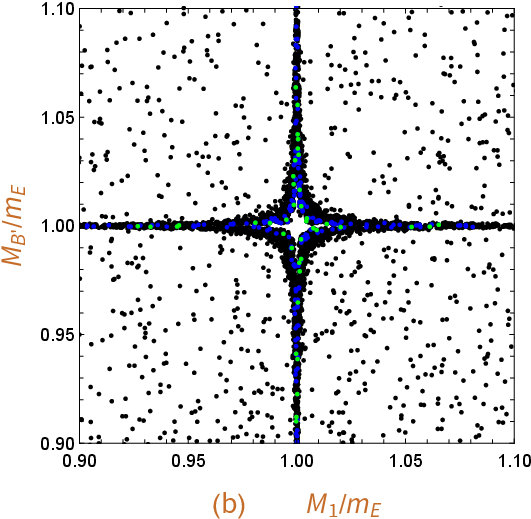}
\caption{Under the premise of limit on CLFV process $\mu\rightarrow e\gamma$, reasonable parameter space is selected to scatter points. Here, the black, blue and green scatter points correspond to the SM-like Higgs boson mass and muon MDM within $4\sigma,3\sigma$ and $2\sigma$ regions, respectively.} \label{fig1}
\end{figure}
\subsection{The CLFV process $\mu\rightarrow e\gamma$}
We all know that the CLFV process $\mu\rightarrow e\gamma$ possesses the strict experimental constraint, so we first discuss this process with $\delta_{13}^{AB}=\delta_{23}^{AB}=0(A,B=L,R),\delta_{12}^{LL}=\delta_{12}^{RR}=\delta_{12}^{LR}=10^{-5}$. Parameters $\tan\beta,M_1,M_2,\mu_H,M_{BB'},M_{B'},A_e,m_L=m_E,g_{YB}$ and $g_B$ are assumed as random variables in the suitable regions. As the SM-like Higgs boson mass and muon MDM are both in $4\sigma,3\sigma$ and $2\sigma$ regions, and the branching ratio of $\mu\rightarrow e\gamma$ satisfies the current experiment constraint, the reasonable parameter space is selected to scatter points, which are shown in FIG.\ref{fig1}. Firstly, we discuss the distribution of $g_B$ versus $g_{YB}$ in FIG.\ref{fig1}(a). Under the experimental constraint of $4\sigma$ region, the numerical results are mainly concentrated in the upper right half interval of function $g_B=-2g_{YB}$. In the $3\sigma$ interval constraint, the numerical results are mostly located in the region above function $g_B=-2g_{YB}$ and below function $g_B=-2g_{YB}+0.35$. In the $2\sigma$ constraint, the numerical results are evenly distributed in the lower left half interval of function $g_B=-2g_{YB}+0.35$. And we find that the region of $g_B$ increases with the decreased $g_{YB}$, even $g_B$ can be anywhere from 0.2 to 0.85 as $g_{YB}\leq-0.25$ with $2\sigma$ limit. So $g_{YB}$ takes $-0.25$ in the following discussion. Besides, in the FIG.\ref{fig1}(b), the parameters $M_1$ and $m_E$ are approximately equal to each other as $\frac{M_{B'}}{m_E}\in(0.9,1.1)$, as well as the parameters $M_{B'}$ and $m_E$ are approximately equal to each other when $\frac{M_{1}}{m_E}\in(0.9,1.1)$. As the experimental constraint changes from $4\sigma$ to $2\sigma$, the values of $\frac{M_{B'}}{m_E}$ and $\frac{M_{1}}{m_E}$ are both closer to 1. Therefore, in the numerical analyses below, we take $M_1\simeq0.947$ TeV, $m_E\simeq0.948$ TeV and $M_{B'}\simeq0.93$ TeV, which are approximately equal to each other. Besides, we appropriately fix $M_2=0.6$ TeV, $\mu_H=0.5$ TeV and $A_e=0.4$ TeV to simplify discussion.
\begin{figure}[t]
\centering
\includegraphics[width=5.3cm]{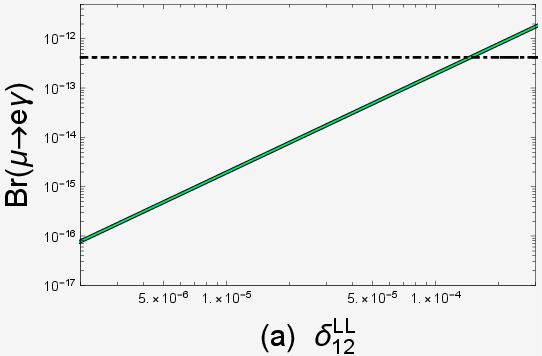}
\includegraphics[width=5.3cm]{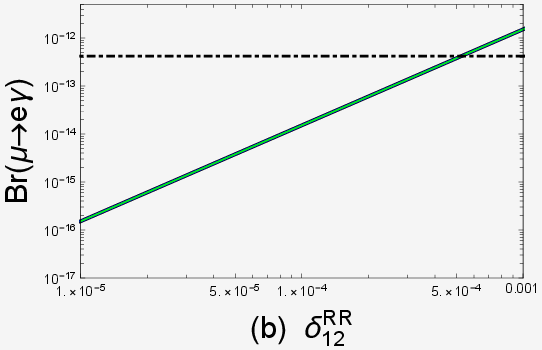}
\includegraphics[width=5.2cm]{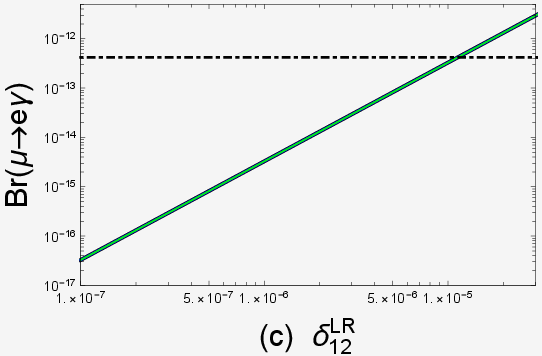}
\caption{The CLFV rates for $\mu\rightarrow e\gamma$ versus $\delta_{12}^{AB}(A,B=L,R)$, where the dotdashed line denotes the present experimental limit. Here, the black, blue and green lines correspond to $4\sigma$, $3\sigma$ and $2\sigma$ regions, respectively.} \label{fig2}
\end{figure}

The CLFV process $\mu\rightarrow e\gamma$ is flavor dependent, which can be influenced by the slepton mixing parameters $\delta_{12}^{AB}(A,B=L,R)$. In order to study the characters of $\delta_{12}^{AB}(A,B=L,R)$ to $Br(\mu\rightarrow e\gamma)$, we assume $g_{YB}=-0.25$, $g_B=0.8$, $M_{BB'}=0.8$ TeV and $\tan\beta=25$, then the corresponding analyses are shown in FIG.\ref{fig2}. It is obvious that the CLFV rates increase with the enlarged $\delta_{12}^{AB}(A,B=L,R)$ and possess the same changes within $4\sigma$, $3\sigma$ and $2\sigma$ constraints. Besides, the present experimental upper bound of $Br(\mu\rightarrow e\gamma)$ constrains $\delta_{12}^{LL}\leq1.4\times10^{-4} $, $\delta_{12}^{RR}\leq 5.1\times10^{-4}$ and $\delta_{12}^{LR}\leq 1.1\times10^{-5}$.
\begin{figure}[t]
\centering
\includegraphics[width=6.6cm]{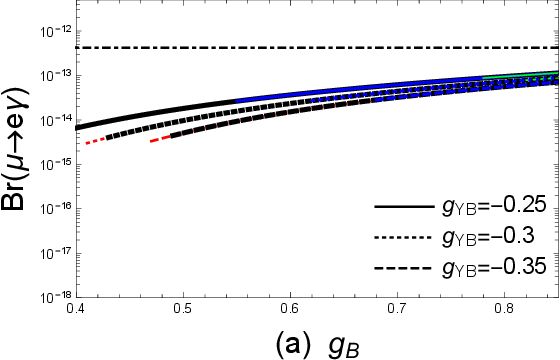}~~~~
\includegraphics[width=6.7cm]{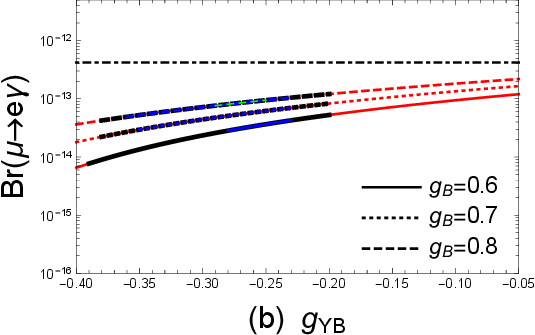}\\
\includegraphics[width=6.6cm]{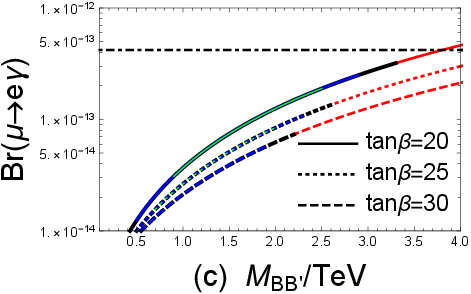}~~~~
\includegraphics[width=6.6cm]{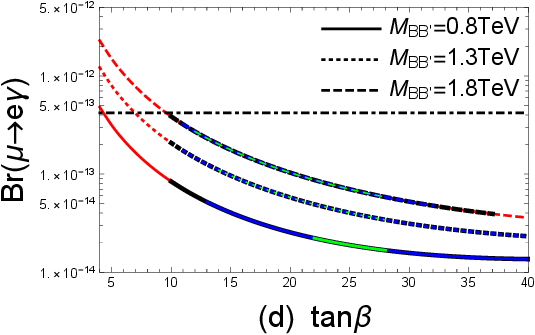}
\caption{The $Br(\mu\rightarrow e\gamma)$ versus $g_B,g_{YB},M_{BB'}$ and $\tan\beta$ respectively, where the dotdashed line denotes the present experimental limit. Here, the black, blue and green lines correspond to $4\sigma,3\sigma$ and $2\sigma$ regions, respectively.} \label{fig3}
\end{figure}

Then, we study the CLFV rates for $\mu\rightarrow e\gamma$ versus $g_B,g_{YB},M_{BB'}$ and $\tan\beta$ respectively in FIG.\ref{fig3}. In general, the numerical results of $Br(\mu\rightarrow e\gamma)$ enlarge with the increase of $g_B,g_{YB},M_{BB'}$, and decrease with the increase of $\tan\beta$. The red line has been excluded due to exceeding $4\sigma$ experimental constraint of SM-like Higgs mass or muon MDM. And the parameter spaces of $g_B,g_{YB},M_{BB'}$ and $\tan\beta$ all narrow with the increase of muon MDM and SM-like Higgs mass constraints. Obviously, FIG.\ref{fig3}(a) and (b) indicate that the larger $g_B$ or $g_{YB}$ is, the more appropriate CLFV rates for $\mu\rightarrow e\gamma$ we can obtain. Under the constraint of $2\sigma$, $g_B$ takes the value in the space of $0.78\sim0.85$ when $g_{YB}=-0.25$, as well as $g_{YB}$ takes the value in the space of $-0.29\sim-0.25$ when $g_{B}=0.8$. In the FIG.\ref{fig3}(c), when the value of $\tan\beta$ is small, the parameter $M_{BB'}$ can obtain a large region, and the upper and lower limits of $M_{BB'}$ also increase. Furthermore, in the $3\sigma$ or $4\sigma$ limit, $\tan\beta$ can obtain a larger region as $M_{BB'}=1.3$ TeV than that of $M_{BB'}=0.8$ TeV or $M_{BB'}=1.8$ TeV. However, in the $2\sigma$ region, the space of $\tan\beta$ decreases with the reduced $M_{BB'}$, and even $\tan\beta$ is confined to a much small interval of $22\sim28$ as $M_{BB'}=0.8$ TeV.
\begin{figure}[t]
\centering
\includegraphics[width=7cm]{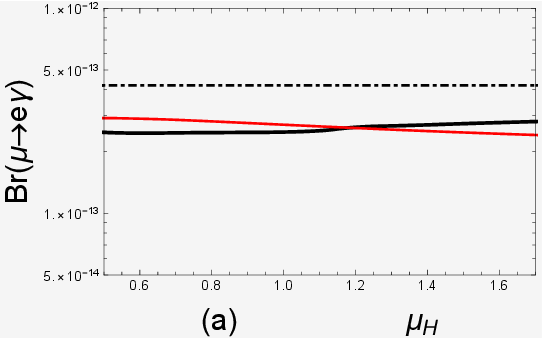}
\hspace{0.8cm}\includegraphics[width=7cm]{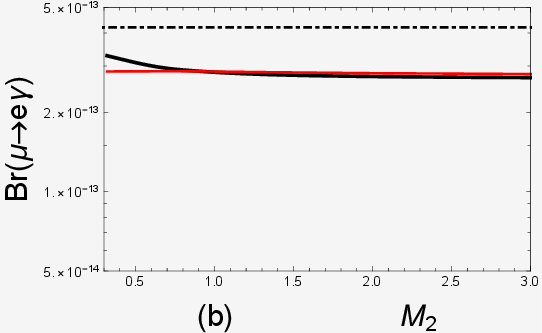}
\caption{The CLFV rates for $\mu\rightarrow e\gamma$ versus $\mu_H$($M_2$), where the black and red lines correspond to the interaction eigenstate and mass eigenstate respectively.} \label{fig6}
\end{figure}

In Refs.\cite{compatibility1,compatibility2}, the authors show that the MIA results can be obtained if one expands the starting expressions in the mass basis properly. The concrete contents include: 1. The loop particles are much heavier than the external states; 2. The convergent of any one-loop amplitude requires that the moduli of every eigenvalue of the dimensionless mass insertion matrix has to be smaller than one. And our discussion satisfies the basic approximation assumption studied in Refs.\cite{compatibility1,compatibility2}. Besides, we further study the numerical results of $Br(\mu \to e \gamma)$ versus $\mu_H$($M_2$) in the interaction eigenstate and mass eigenstate respectively in FIG.\ref{fig6}. When parameters are valued in a reasonable space, such as $\delta_{12}^{LL}=5\times10^{-5}$, $\delta_{12}^{RR}=5\times10^{-4}$, $\delta_{12}^{LR}=9.5\times10^{-6}$, $g_{YB}=-0.25$, $g_B=0.8$ and $\tan\beta=15$, we find that the $Br(\mu \to e \gamma)$ with the running of $\mu_H$($M_2$) in the interaction eigenstate have the slight deviation comparing with that in mass eigenstate. Even when $\mu_H$ or $M_2$ takes a certain value, the branch ratios in interaction eigenstate and mass eigenstate are exactly the same. In this work, the numerical discussion of MIA scheme in the interaction eigenstates is based on satisfying the numerical results of mass eigenstates. Therefore, the compatibility between SARAH and the MIA scheme is guaranteed in this work.
\subsection{The CLFV processes $\tau\rightarrow e\gamma$ and $\tau\rightarrow \mu\gamma$}
\begin{figure}[t]
\centering
\includegraphics[width=5.3cm]{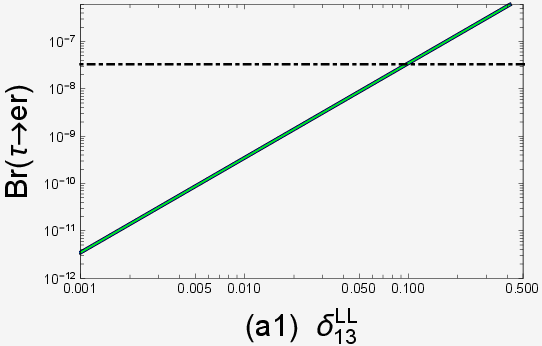}
\includegraphics[width=5.3cm]{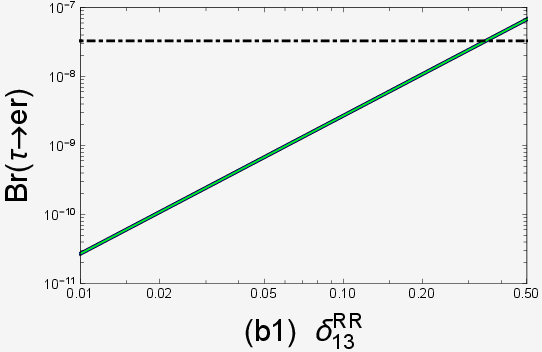}
\includegraphics[width=5.1cm]{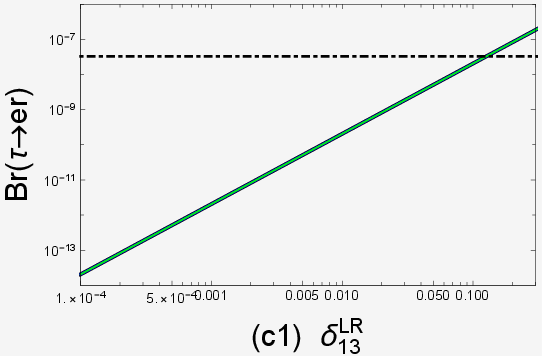}\\
\includegraphics[width=5.3cm]{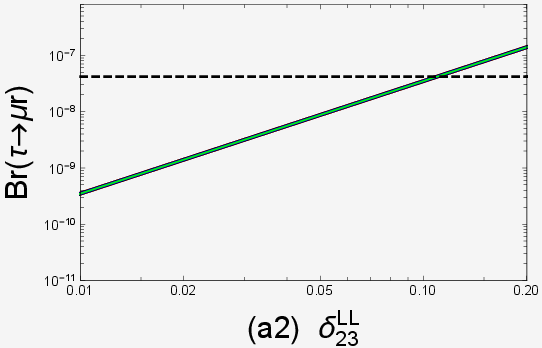}
\includegraphics[width=5.3cm]{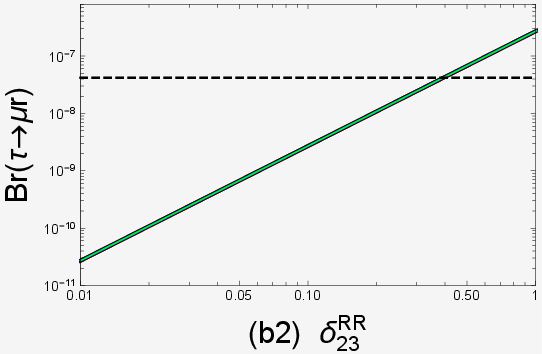}
\includegraphics[width=5.3cm]{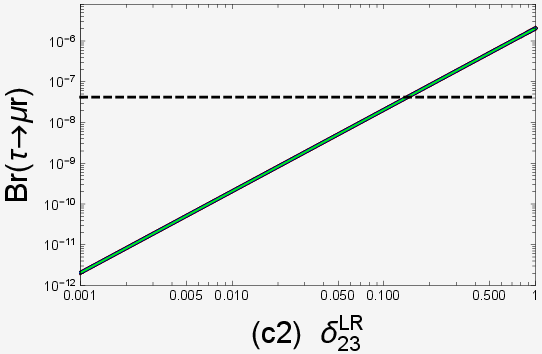}
\caption{The CLFV rates for $\tau\rightarrow e\gamma (\tau\rightarrow \mu\gamma)$ versus $\delta_{13}^{AB}(\delta_{23}^{AB}(A,B=L,R))$, where the dotdashed lines denote the present experimental limits. Here, the black, blue and green lines correspond to $4\sigma,3\sigma$ and $2\sigma$ regions, respectively.} \label{fig4}
\end{figure}

Due that the experimental upper bounds for $Br(\tau\rightarrow e\gamma)$ and $Br(\tau\rightarrow \mu\gamma)$ do not possess obvious difference, so we research both these processes in this section. In FIG.\ref{fig4}, we picture $Br(\tau\rightarrow e\gamma)$ ($Br(\tau\rightarrow \mu\gamma)$) versus the slepton flavor mixing parameters $\delta_{13}^{AB}(A,B=L,R)$ ($\delta_{23}^{AB}(A,B=L,R)$). Obviously, the CLFV rates of $\tau\rightarrow e\gamma$ possess the same trends for $4\sigma$, $3\sigma$ and $2\sigma$ limits, as well as $Br(\tau\rightarrow \mu\gamma)$. We can clearly see that all slepton flavor mixing parameters take positive influences on the CLFV rates, even the $Br(\tau\rightarrow e\gamma)$ and $Br(\tau\rightarrow \mu\gamma)$ are respectively proportional to $\delta_{13}^{AB}$ and $\delta_{23}^{AB}(A,B=L,R)$. Besides, we find that $\delta_{13}^{LL}\leq0.09 $, $\delta_{13}^{RR}\leq 0.35$ and $\delta_{13}^{LR}\leq 0.13$ with the experimental constraint of $Br(\tau\rightarrow e\gamma)$. Through the present limit of $Br(\tau\rightarrow \mu\gamma)$, $\delta_{23}^{LL}$, $\delta_{23}^{RR}$ and $\delta_{23}^{LR}$ are restricted below 0.11, 0.39 and 0.14 respectively.
\begin{figure}[t]
\centering
\includegraphics[width=6.3cm]{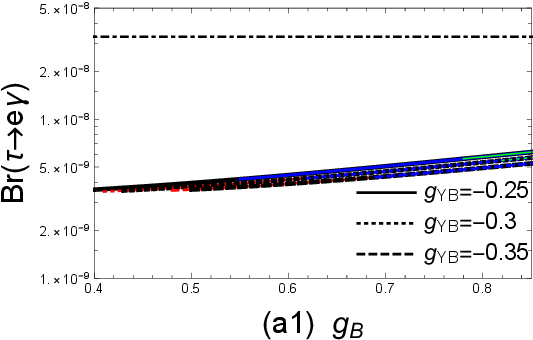}~~~~
\includegraphics[width=6.4cm]{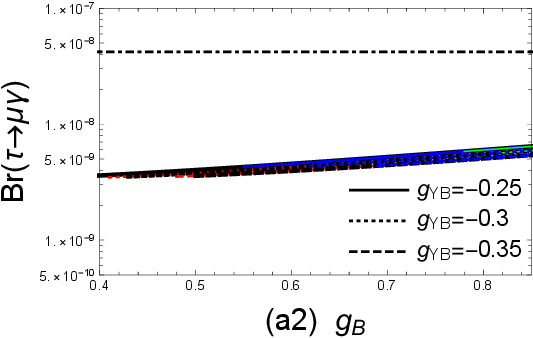}\\
\includegraphics[width=6.5cm]{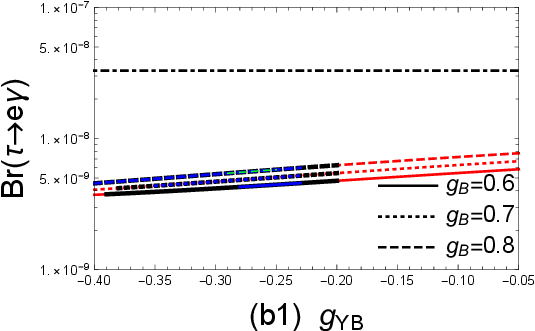}~~~~
\includegraphics[width=6.5cm]{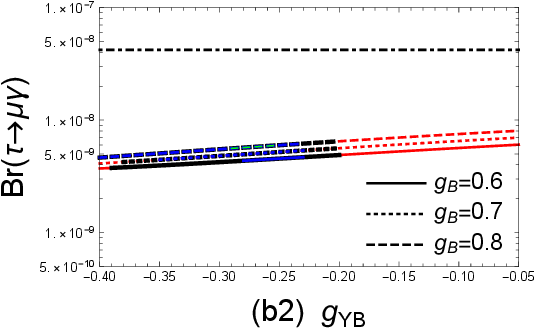}\\
\includegraphics[width=6.4cm]{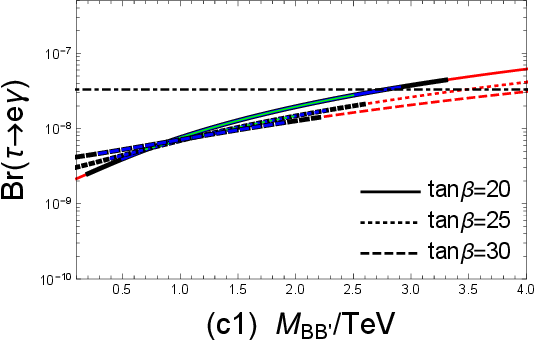}~~~~
\includegraphics[width=6.4cm]{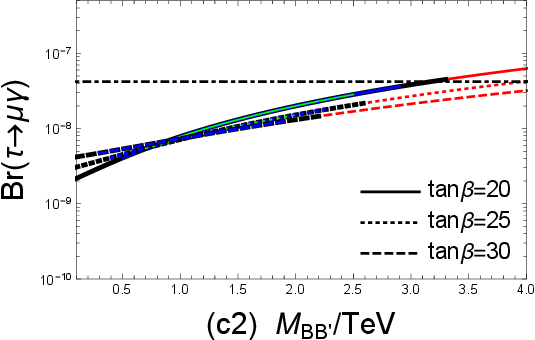}\\
\includegraphics[width=6.5cm]{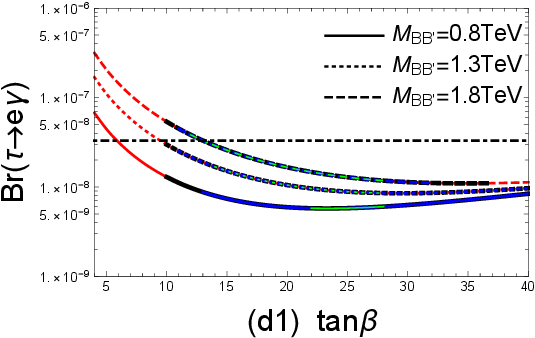}~~~~
\includegraphics[width=6.5cm]{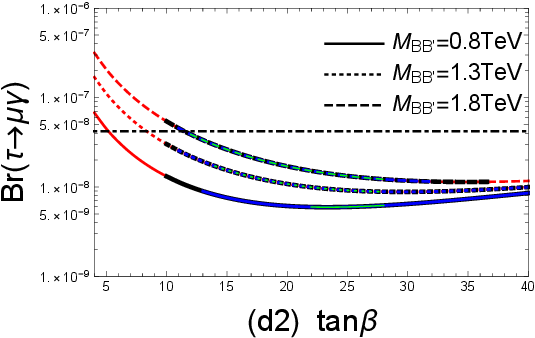}\\
\caption{The CLFV rates change with $g_B,g_{YB},M_{BB'}$ and $\tan\beta$, where the black, blue and green lines correspond to $4\sigma,3\sigma$ and $2\sigma$ regions, and the dotdashed line denotes the experimental limit.} \label{fig5}
\end{figure}

In order to discuss the effects from other basic parameters on the numerical results, we set $\delta_{13}^{LL}=\delta_{13}^{RR}=\delta_{13}^{LR}=0.03$ and $\delta_{23}^{AB}(A,B=L,R)=0$ for process $\tau\rightarrow e\gamma$, as well as $\delta_{23}^{LL}=\delta_{23}^{RR}=\delta_{23}^{LR}=0.03$ and $\delta_{13}^{AB}(A,B=L,R)=0$ for process $\tau\rightarrow \mu\gamma$. Then, we study the influences on $Br(\tau\rightarrow e\gamma)$ and $Br(\tau\rightarrow \mu\gamma)$ from parameters $g_B,g_{YB},M_{BB'}$ and $\tan\beta$. Overall, the variations of $Br(\tau\rightarrow e\gamma)$ with parameters $g_B,g_{YB},M_{BB'}$ and $\tan\beta$ are all slightly steeper than that of $Br(\tau\rightarrow \mu\gamma)$, and the value spaces of $g_B,g_{YB},M_{BB'}$ and $\tan\beta$ become small as the constraints of SM-like Higgs mass and muon MDM change from $4\sigma$ to $2\sigma$. FIG.\ref{fig5}(a1) and (a2) both indicate that the ranges of $g_B$ increase as $g_{YB}$ expands, and when $g_{YB}=-0.25$, $g_B$ can fetch all values from 0.78 to 0.85 under the $2\sigma$ constraint. $g_{YB}$ in FIG.\ref{fig5}(b1) and (b2) can both obtain smallest regions when $g_B=0.7$ with the limits of $4\sigma$. However, within $3\sigma$ limit, the larger $g_B$ is, the larger space for reasonable values $g_{YB}$ is. Under $2\sigma$ constraint, $g_{YB}$ runs in a $-0.29\sim-0.25$ cell only when $g_B=0.8$.

In FIG.\ref{fig5}(c1) and (c2), we can see that the large regions of $M_{BB'}$ have been eliminated by the experimental upper limits of $Br(\tau\rightarrow e\gamma)$ and $Br(\tau\rightarrow \mu\gamma)$ both within $4\sigma$ or $3\sigma$ limit, but $M_{BB'}$ can still obtain large region as $\tan\beta=20$. Besides, the experimental limit of $Br(\tau\rightarrow e\gamma)$ is stronger than that of $Br(\tau\rightarrow \mu\gamma)$. Under the constraint of $2\sigma$, the value range of $M_{BB'}$ changes from 0.7 TeV to 2 TeV as $\tan\beta=25$, which is more narrow than that of $\tan\beta=20$ (here 0.9 TeV$\leq M_{BB'}\leq$2.5 TeV). In addition, considering $4\sigma$ or $3\sigma$ constraint on the FIG.\ref{fig5}(d1) and (d2), we find that the larger $M_{BB'}$ is, the more obvious experimental constraints of $Br(\tau\rightarrow e\gamma)$ and $Br(\tau\rightarrow \mu\gamma)$ on $\tan\beta$ are, and the smaller value space of $\tan\beta$ is. That is to say the suitable parameter space of $\tan\beta$ narrows obviously as $M_{BB'}$ enlarges. When $M_{BB'}=1.8$ TeV, $\tan\beta$ can be any value between 14 and 28 within $2\sigma$ region, which indicates parameter $\tan\beta$ is more constrained by the experimental limit of $Br(\tau\rightarrow e\gamma)$ than $Br(\tau\rightarrow \mu\gamma)$.
\section{discussion and conclusion}
In this work, under the premise that the compatibility between SARAH and the MIA scheme is guaranteed, we focus on CLFV processes $l_j^-\rightarrow l_i^- \gamma$ in the B-LSSM using the MIA method. Assuming that all superparticle masses are almost degenerate, we find that the B-LSSM contributions for both muon MDM and CLFV processes beyond MSSM are considerable at the analytical level. In the numerical discussion, we constrain the SM-like Higgs mass and muon MDM both within $4\sigma$, $3\sigma$ and $2\sigma$ regions, which possess strict limits on the CLFV decays. Firstly, the distribution of $g_B$ versus $g_{YB}$ presents different sensitive parameter regions under the different bounds within $2\sigma$, $3\sigma$ and $4\sigma$. Besides, the parameters $M_1$ and $m_E$ are approximately equal to each other as $\frac{M_{B'}}{m_E}\in(0.9,1.1)$, as well as the parameters $M_{B'}$ and $m_E$ are approximately equal to each other when $\frac{M_{1}}{m_E}\in(0.9,1.1)$. As the experimental constraint changes from $4\sigma$ to $2\sigma$, the values of $\frac{M_{B'}}{m_E}$ and $\frac{M_{1}}{m_E}$ are both closer to 1. The branching ratios of $l_j^-\rightarrow l_i^- \gamma$ depend on the slepton flavor mixing parameters $\delta_{12}^{AB},\delta_{13}^{AB}$ and $\delta_{23}^{AB}(A,B=L,R)$ obviously, and $Br(l_j^-\rightarrow l_i^- \gamma)$ increase with the enlarged $\delta_{12}^{AB},\delta_{13}^{AB}$ and $\delta_{23}^{AB}$. Considering the latest experimental limits of $Br(l_j^-\rightarrow l_i^- \gamma)$, the slepton flavor mixing parameters are restricted as $\delta_{12}^{LL}\leq1.4\times10^{-4} $, $\delta_{12}^{RR}\leq 5.1\times10^{-4}$, $\delta_{12}^{LR}\leq 1.1 \times10^{-5}$, $\delta_{13}^{LL}\leq0.09 $, $\delta_{13}^{RR}\leq 0.35$, $\delta_{13}^{LR}\leq 0.13$, $\delta_{23}^{LL}\leq0.11$, $\delta_{23}^{RR}\leq0.39$ and $\delta_{23}^{LR}\leq 0.14$.

The parameter ranges of $g_B$, $g_{YB}$, $M_{BB'}$ and $\tan\beta$ all narrow obviously as the constraints of SM-like Higgs mass and muon MDM change from $4\sigma$ to $2\sigma$. Compared with the MSSM, $g_B$, $g_{YB}$ and $M_{BB'}$ are the new parameters in the B-LSSM, which have an uplift effects on the numerical results. At $3\sigma$ or $2\sigma$ limit, $g_B$ has a large region with the increased $g_{YB}$, similarly, the reasonable parameter space of $g_{YB}$ also widens as the value of $g_B$ enlarges. When $g_{YB}=-0.25$($g_B=0.8$), $g_B$($g_{YB}$) is constrained in the range of $0.78\sim0.85$($-0.29\sim-0.25$), which correspond to the maximum spaces of these parameters under the $2\sigma$ constraint. Under the $2\sigma$ constraint, the parameter $M_{BB'}$ can obtain a large reasonable space $0.9\sim2.5$ TeV when $\tan\beta=20$. The large parameter space of $\tan\beta$ is excluded by the latest experiment limit of $Br(\tau\rightarrow e\gamma)$, which possesses the strongest constraint. Therefore, under the $2\sigma$ and $Br(\tau\rightarrow e\gamma)$ constraints, $\tan\beta$ can run in a much large region of $14\sim28$ when $M_{BB'}=1.8$ TeV.

{\bf Acknowledgments}

This work is supported by the Major Project of National Natural Science Foundation of China (NNSFC) (No. 12235008), the National Natural Science Foundation of China (NNSFC) (No. 12075074, No. 12075073), the Natural Science Foundation of Hebei province(No. A202201022, No. A2020201002), the Natural Science Foundation of Hebei Education Department(No. QN2022173).

\appendix
\section{One-loop functions}
The one-loop functions $\mathcal{I}(x, y, z)$, $\mathcal{J}(x, y, z)$, $f(x, y, z, t)$, $g(x, y, z, t)$ and $k(x, y, z, t)$ can be written as
\begin{eqnarray}
&&\mathcal{I}(x, y, z)=\frac{1}{16 \pi^{2}}\Big[\frac{1}{(x-z)(z-y)}+\frac{(z^{2}-x y) \log z}{(x-z)^{2}(y-z)^{2}}-\frac{x \log x}{(x-y)(x-z)^{2}}\nonumber\\
&&\hspace{2.5cm}+\frac{y \log y}{(x-y)(y-z)^{2}}\Big], \nonumber\\
&& \mathcal{J}(x, y, z)=\frac{1}{16 \pi^{2}}\Big[\frac{x(x^{2}+x z-2 y z) \log x}{(x-y)^{2}(x-z)^{3}}-\frac{y^{2} \log y}{(x-y)^{2}(y-z)^{2}}+\frac{z[x(z-2 y)+z^{2}] \log z}{(z-x)^{3}(y-z)^{2}}\nonumber\\
&&\hspace{2.5cm}-\frac{x(y-2 z)+y z}{(x-y)(x-z)^{2}(y-z)}\Big],\nonumber\\
&&f(x, y, z, t)=\frac{1}{16 \pi^{2}}\Big[\frac{t\left[t^{3}-3 t x y+x y(x+y)\right] \log t}{(t-x)^{3}(t-y)^{3}(t-z)}-\frac{x\left[x^{3}-3 t x z+t z(t+z)\right] \log x}{(t-x)^{3}(x-y)(x-z)^{3}}\nonumber\\
&&\hspace{1cm}+\frac{y\left[y^{3}-3 t y z+t z(t+z)\right] \log y}{(t-y)^{3}(x-y)(y-z)^{3}}-\frac{z\left[z^{3}-3 x y z+x y(x+y)\right] \log z}{(t-z)(z-x)^{3}(z-y)^{3}}+\frac{1}{2(x-y)} \nonumber\\
&&\hspace{1cm}\times\left(\frac{t}{(t-x)^{2}(z-x)}-\frac{2 y}{(t-y)(y-z)^{2}}+\frac{x(2 t-3 x+z)}{(t-x)^{2}(x-z)^{2}}+\frac{t+y}{(t-y)^{2}(y-z)}\right)\Big],\nonumber\\
&&g(x, y, z, t)=\frac{1}{16 \pi^{2}}\Big\{-\frac{t\left[t^{3}(x+y)-3 t^{2} x y+x^{2} y^{2}\right] \log t}{(t-x)^{3}(t-y)^{3}(t-z)}\nonumber\\
&&\hspace{1cm}+\frac{z\left[x^{2} y^{2}+x z^{2}(z-3 y)+y z^{3}\right] \log z}{(t-z)(z-x)^{3}(z-y)^{3}}+\frac{x^{2}\left[x^{3}-3 t x z+t z(t+z)\right] \log x}{(t-x)^{3}(x-y)(x-z)^{3}} \nonumber\\
&&\hspace{1cm}-\frac{y^{2}\left[y^{3}-3 t y z+t z(t+z)\right] \log y}{(t-y)^{3}(x-y)(y-z)^{3}}-\frac{x^{2}(2 t-3 x+z)}{2(t-x)^{2}(x-y)(x-z)^{2}} \nonumber\\
&&\hspace{1cm}+\frac{t x}{2(t-x)^{2}(x-y)(x-z)}-\frac{y[t(y+z)+y(z-3 y)]}{2(t-y)^{2}(y-x)(y-z)^{2}}\Big\},\nonumber\\
&&k(x, y, z, t) =\frac{1}{16 \pi^{2}}\Big\{-\frac{\left[z \left(x^2-t y\right)+x^2 (t-2x+y)\right]\log x }{(t-x)^2 (x-y)^2 (x-z)^2}-\frac{t \log t}{(t-x)^2 (t-y) (t-z)}\nonumber\\
&&\hspace{1cm}+\frac{y \log y}{(t-y)
   (x-y)^2 (y-z)}+\frac{1}{(t-x)
   (x-y)(x-z)}+\frac{z \log z}{(t-z) (z-y)(x-z)^2}\Big\}.
\end{eqnarray}

\section{The chirality flips}
The chirality flips occurring in the internal gaugino and the external lepton lines were described in detail in the Refs.\cite{chirality flip1,chirality flip2,chirality flip3}. Since the gaugino mass is much bigger than the lepton mass, these diagrams whose chirality flips occur in the internal gaugino lines may yield dominant contributions.
\begin{figure}[t]
\centering
\includegraphics[width=12cm]{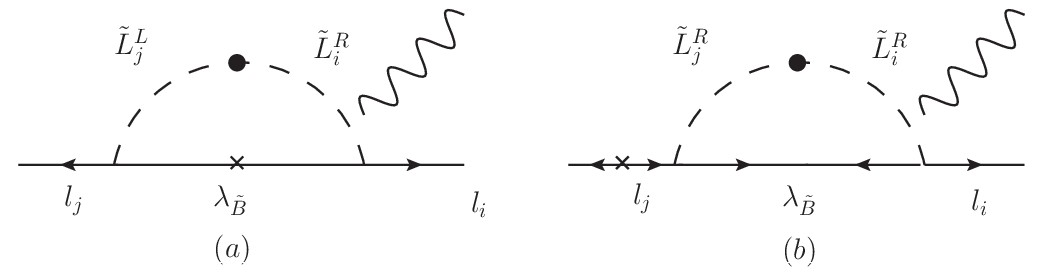}
\caption{The feynman diagrams of $l_j^-\rightarrow l_i^- \gamma$ by MIA in the B-LSSM, the chirality flips occur in the internal gaugino and external lepton lines, respectively.} \label{fig7}
\end{figure}

Then, we take two Feynman pictures shown in FIG.\ref{fig7} as the examples, and discuss the chirality flips occurring in the internal gaugino and the external lepton lines as follows. Firstly, we discuss the concrete amplitude of FIG.\ref{fig7}(a):
\begin{eqnarray}
&&i{\cal M}_1\hspace{-0.1cm}=\hspace{-0.1cm}\bar{u}_i(p+q)
\mu^{4-D}\hspace{-0.2cm}\int\hspace{-0.2cm}\frac{d^Dk}{(2\pi)^D}\hspace{-0.1cm}\Big[\frac{eA_LP_L({/\!\!\!k}+M_1)B_LP_L(2p+q-2k)^{\mu}C}
{[k^2\hspace{-0.1cm}-\hspace{-0.1cm}M_1^2][(k\hspace{-0.1cm}-\hspace{-0.1cm}p)^2\hspace{-0.1cm}-\hspace{-0.1cm}m_{\tilde{L}_j^L}^2][(k\hspace{-0.1cm}-\hspace{-0.1cm}(p\hspace{-0.1cm}+\hspace{-0.1cm}q))^2\hspace{-0.1cm}-\hspace{-0.1cm}m_{\tilde{L}_j^L}^2][(k\hspace{-0.1cm}-\hspace{-0.1cm}(p\hspace{-0.1cm}+\hspace{-0.1cm}q))^2\hspace{-0.1cm}-\hspace{-0.1cm}m_{\tilde{L}_i^{R}}^2]}
\nonumber\\&&\hspace{0.8cm}+\frac{eA_LP_L({/\!\!\!k}+M_1)B_LP_L(2p+q-2k)^{\mu}C}
{[k^2-M_1^2][(k-p)^2-m_{\tilde{L}_j^L}^2][(k-p)^2-m_{\tilde{L}_i^{R}}^2][(k-(p+q))^2-m_{\tilde{L}_i^{R}}^2]}\Big]u_j(p)\epsilon_{\mu}(q).
\end{eqnarray}
We assume that the internal masses of particles such as gaugino and left(right)-handed slepton are much bigger than the external lepton mass, the function $\frac{1}{(k-p)^2-m^2}\simeq\frac{1}{k^2-m^2}+\frac{2k\cdot p-p^2}{(k^2-m^2)^2}+\frac{4(k\cdot p)^2}{(k^2-m^2)3}$. Then the amplitude can be approximately reduced as
\begin{eqnarray}
&&i{\cal M}_1\simeq\bar{u}_i(p+q)
eM_1CA_{L}B_{L}P_L\mu^{4-D}\int\frac{d^Dk}{(2\pi)^D}\frac{(2p+q)^{\mu}}
{[k^2-M_1^2][k^2-m_{\tilde{L}_j^L}^2][k^2-m_{\tilde{L}_i^{R}}^2]}\Big[\frac{1}{k^2-m_{\tilde{L}_j^L}^2}
\nonumber\\&&+\frac{1}{k^2-m_{\tilde{L}_i^{R}}^2}
-\frac{k^2}{(k^2-m_{\tilde{L}_j^L}^2)^2}-\frac{k^2}{(k^2-m_{\tilde{L}_i^{R}}^2)^2}-\frac{k^2}{(k^2-m_{\tilde{L}_j^L}^2)(k^2-m_{\tilde{L}_i^{R}}^2)}
\Big]u_j(p)\epsilon_{\mu}(q)
\nonumber\\&&=-\bar{u}_i(p+q)eM_1CA_{L}B_{L}P_Li\sigma^{\mu\nu}q_{\nu}
\frac{i}{2\Lambda^4}\Big[\mathcal{J}(x_{1}, x_{\tilde{L}_j^L},x_{\tilde{L}_i^{R}})\hspace{-0.1cm}+\hspace{-0.1cm}\mathcal{J}(x_{1}, x_{\tilde{L}_i^{R}}, x_{\tilde{L}_j^L})\Big]u_j(p)\epsilon_{\mu}(q).
\end{eqnarray}
With $A_L=-\sqrt{2}g_1$, $B_L=\frac{g_1}{\sqrt{2}}$ and $C=-\bigtriangleup_{ij}^{LR}=-A_e v_d \delta_{ij}^{LR}$, the Wilson coefficient of FIG.\ref{fig7}(a) can be written as:
\begin{eqnarray}
&&C_2\left(\tilde{L}_j^L,\tilde{L}_i^{R}, \lambda_{\tilde{B}}\right)=-\frac{g_{1}^{2}} {2\Lambda^4 m_{l_j}} M_1A_e v_d \delta_{ij}^{LR}P_L \Big[\mathcal{J}(x_{1}, x_{\tilde{L}_j^L},x_{\tilde{L}_i^{R}})+\mathcal{J}(x_{1}, x_{\tilde{L}_i^{R}}, x_{\tilde{L}_j^L})\Big].
\end{eqnarray}
Here, the chirality flip occurs in the internal gaugino line.

Secondly, we take FIG.\ref{fig7}(b) as an example, and discuss the chirality flip occurs in the external lepton line. The concrete amplitude is written as:
\begin{eqnarray}
&&i{\cal M}_2\hspace{-0.1cm}=\hspace{-0.1cm}\bar{u}_i(p+q)
\mu^{4-D}\hspace{-0.2cm}\int\hspace{-0.2cm}\frac{d^Dk}{(2\pi)^D}\hspace{-0.1cm}\Big[\frac{eA_LP_L({/\!\!\!k}+M_1)B_RP_R(2p+q-2k)^{\mu}C'}
{[k^2\hspace{-0.1cm}-\hspace{-0.1cm}M_1^2][(k\hspace{-0.1cm}-\hspace{-0.1cm}p)^2\hspace{-0.1cm}-\hspace{-0.1cm}m_{\tilde{L}_j^R}^2][(k\hspace{-0.1cm}-\hspace{-0.1cm}(p\hspace{-0.1cm}+\hspace{-0.1cm}q))^2\hspace{-0.1cm}-\hspace{-0.1cm}m_{\tilde{L}_j^R}^2][(k\hspace{-0.1cm}-\hspace{-0.1cm}(p\hspace{-0.1cm}+\hspace{-0.1cm}q))^2\hspace{-0.1cm}-\hspace{-0.1cm}m_{\tilde{L}_i^{R}}^2]}
\nonumber\\&&\hspace{0.8cm}+\frac{eA_LP_L({/\!\!\!k}+M_1)B_RP_R(2p+q-2k)^{\mu}C'}
{[k^2-M_1^2][(k-p)^2-m_{\tilde{L}_j^R}^2][(k-p)^2-m_{\tilde{L}_i^{R}}^2][(k-(p+q))^2-m_{\tilde{L}_i^{R}}^2]}\Big]u_j(p)\epsilon_{\mu}(q)
\nonumber\\&&\hspace{0.8cm}\simeq\bar{u}_i(p+q)
eC'A_{L}B_{R}P_L\mu^{4-D}\int\frac{d^Dk}{(2\pi)^D}\frac{{/\!\!\!k}(2p+q)^{\mu}2k\cdot(2p+q)}
{[k^2-M_1^2][k^2-m_{\tilde{L}_j^R}^2][k^2-m_{\tilde{L}_i^{R}}^2]}\Big[\frac{1}{(k^2-m_{\tilde{L}_j^R}^2)^2}
\nonumber\\&&\hspace{0.8cm}+\frac{1}{(k^2-m_{\tilde{L}_j^R}^2)(k^2-m_{\tilde{L}_i^{R}}^2)}
+\frac{1}{(k^2-m_{\tilde{L}_i^{R}}^2)^2}
\Big]u_j(p)\epsilon_{\mu}(q)
\nonumber\\&&\hspace{0.8cm}\simeq\bar{u}_i(p+q)
eC'A_{L}B_{R}P_L\mu^{4-D}\int\frac{d^Dk}{(2\pi)^D}\frac{1}{2}\frac{(2p+q)^{\mu}(2{/\!\!\!p}+{/\!\!\!q})}
{[k^2-M_1^2][k^2-m_{\tilde{L}_j^R}^2][k^2-m_{\tilde{L}_i^{R}}^2]}\Big[\frac{k^2}{(k^2-m_{\tilde{L}_j^R}^2)^2}
\nonumber\\&&\hspace{0.8cm}+\frac{k^2}{(k^2-m_{\tilde{L}_j^R}^2)(k^2-m_{\tilde{L}_i^{R}}^2)}
+\frac{k^2}{(k^2-m_{\tilde{L}_i^{R}}^2)^2}
\Big]u_j(p)\epsilon_{\mu}(q)
\nonumber\\&&\hspace{0.8cm}=-\bar{u}_i(p+q)eC'A_{L}B_{R}(\frac{m_{l_j}}{2}P_L+\frac{m_{l_i}}{2}P_R)i\sigma^{\mu\nu}q_{\nu}
\frac{i}{2\Lambda^4}\Big[2\mathcal{I}( x_{\tilde{L}_j^R},x_{1},x_{\tilde{L}_i^{R}})
\nonumber\\&&\hspace{0.8cm}+2\mathcal{I}( x_{\tilde{L}_i^{R}},x_{1}, x_{\tilde{L}_j^R})-\mathcal{J}( x_{\tilde{L}_j^R},x_{\tilde{L}_i^{R}},x_{1})\hspace{-0.1cm}-\hspace{-0.1cm}\mathcal{J}( x_{\tilde{L}_i^{R}}, x_{\tilde{L}_j^R},x_{1})\Big]u_j(p)\epsilon_{\mu}(q).
\end{eqnarray}
With $A_L=B_R =-\sqrt{2}g_1$, $C'=-\bigtriangleup_{ij}^{RR}=-m_{{E}}^2\delta_{ij}^{RR}$ and $m_{l_j}\gg m_{l_i}$, the Wilson coefficient of FIG.\ref{fig7}(b) can be written as:
\begin{eqnarray}
&&C_2'\left(\tilde{L}_j^R,\tilde{L}_i^{R}, \lambda_{\tilde{B}}\right)\simeq\frac{g_{1}^{2}} {2\Lambda^4}m_{{E}}^2\delta_{ij}^{RR} P_L \Big[2\mathcal{I}( x_{\tilde{L}_j^R},x_{1},x_{\tilde{L}_i^{R}})+2\mathcal{I}( x_{\tilde{L}_i^{R}},x_{1}, x_{\tilde{L}_j^R})
\nonumber\\&&\hspace{3cm}-\mathcal{J}( x_{\tilde{L}_j^R},x_{\tilde{L}_i^{R}},x_{1})\hspace{-0.1cm}-\hspace{-0.1cm}\mathcal{J}( x_{\tilde{L}_i^{R}}, x_{\tilde{L}_j^R},x_{1})\Big].
\end{eqnarray}

If $|M_{1}|=m_{{L}}=m_{{E}}=A_e=\Lambda$, then $C_2\left(\tilde{L}_j^L,\tilde{L}_i^{R}, \lambda_{\tilde{B}}\right)\rightarrow -\frac{g_{1}^{2}} {192\pi^2\Lambda^4 m_{l_j}} M_1A_e v_d \delta_{ij}^{LR}P_L$, and $C_2'\left(\tilde{L}_j^R,\tilde{L}_i^{R}, \lambda_{\tilde{B}}\right)\rightarrow\frac{g_{1}^{2}} {64 \pi^2 \Lambda^4 m_{l_j}} m_{l_j}m_{{E}}^2\delta_{ij}^{RR}P_L $. If $\delta_{ij}^{RR}$ and $\delta_{ij}^{LR}$ in the same order, $M_1A_e v_d\gg m_{l_j}m_{{E}}^2$, then the absolute value of $C_2\left(\tilde{L}_j^L,\tilde{L}_i^{R}, \lambda_{\tilde{B}}\right)$ is much bigger than $C_2'\left(\tilde{L}_j^R,\tilde{L}_i^{R}, \lambda_{\tilde{B}}\right)$. Therefore, we omit the chirality flip of incident lepton in our calculation.

\section{The comparisons from FIG.\ref{figb}(a1) and (a2) }
We will discuss the one-loop contributions from $\tilde{H}^{-}-\tilde{W}^{-}-\tilde{\nu}_{L}^{R}$ in detail, which derive from FIG.\ref{figb}(a1). The concrete amplitude is
\begin{eqnarray}
&&\hspace{-2cm}i{\cal M}_{a1}\hspace{-0.1cm}=\hspace{-0.1cm}\bar{u}_i(p+q)
\hspace{-0.2cm}\int\hspace{-0.2cm}\frac{d^4k}{(2\pi)^D}\hspace{-0.1cm}\Big[\frac{eA_LP_L({/\!\!\!k\hspace{-0.1cm}+\hspace{-0.1cm}/\!\!\!p\hspace{-0.1cm}+\hspace{-0.1cm}/\!\!\!q}\hspace{-0.1cm}+\hspace{-0.1cm}\mu_H)(B_LP_L\hspace{-0.1cm}+\hspace{-0.1cm}B_RP_R)({/\!\!\!k\hspace{-0.1cm}+\hspace{-0.1cm}/\!\!\!p\hspace{-0.1cm}+\hspace{-0.1cm}/\!\!\!q}\hspace{-0.1cm}+\hspace{-0.1cm}M_2)\gamma^{\mu}({/\!\!\!k\hspace{-0.1cm}+\hspace{-0.1cm}/\!\!\!p}\hspace{-0.1cm}+\hspace{-0.1cm}M_2)C_LP_LD}
{[(k+p+q)^2\hspace{-0.1cm}-\hspace{-0.1cm}\mu_H^2][(k+p+q)^2\hspace{-0.1cm}-\hspace{-0.1cm}M_2^2][(k+p)^2\hspace{-0.1cm}-\hspace{-0.1cm}M_2^2][k^2\hspace{-0.1cm}-\hspace{-0.1cm}m_{\tilde{\nu}_{L_j}^{R}}^2][k^2\hspace{-0.1cm}-\hspace{-0.1cm}m_{\tilde{\nu}_{L_i}^{R}}^2]}
\nonumber\\&&\hspace{0cm}+\frac{eA_LP_L({/\!\!\!k\hspace{-0.1cm}+\hspace{-0.1cm}/\!\!\!p\hspace{-0.1cm}+\hspace{-0.1cm}/\!\!\!q}\hspace{-0.1cm}+\hspace{-0.1cm}\mu_H)\gamma^{\mu}({/\!\!\!k\hspace{-0.1cm}+\hspace{-0.1cm}/\!\!\!p}\hspace{-0.1cm}+\hspace{-0.1cm}\mu_H)(B_LP_L\hspace{-0.1cm}+\hspace{-0.1cm}B_RP_R)({/\!\!\!k\hspace{-0.1cm}+\hspace{-0.1cm}/\!\!\!p}\hspace{-0.1cm}+\hspace{-0.1cm}M_2)C_LP_LD}
{[(k+p+q)^2\hspace{-0.1cm}-\hspace{-0.1cm}\mu_H^2][(k+p)^2\hspace{-0.1cm}-\hspace{-0.1cm}\mu_H^2][(k+p)^2\hspace{-0.1cm}-\hspace{-0.1cm}M_2^2][k^2\hspace{-0.1cm}-\hspace{-0.1cm}m_{\tilde{\nu}_{L_j}^{R}}^2][k^2\hspace{-0.1cm}-\hspace{-0.1cm}m_{\tilde{\nu}_{L_i}^{R}}^2]}
\Big]u_j(p)\epsilon_{\mu}(q).
\end{eqnarray}
We assume that the internal masses of particles are much bigger than the external lepton mass, the function $\frac{1}{(k+p)^2-m^2}\simeq\frac{1}{k^2-m^2}-\frac{2k\cdot p+p^2}{(k^2-m^2)^2}+\frac{4(k\cdot p)^2}{(k^2-m^2)3}$. With $A_L=\frac{Y_{l_i}}{\sqrt{2}}$, $B_L=-\frac{g_2}{\sqrt{2}}v_d \tan\beta$, $B_R=-\frac{g_2}{\sqrt{2}}v_d$ (here $B_L\gg B_R$), $C_L=\frac{g_2}{\sqrt{2}}$ and $D=-\bigtriangleup_{ij}^{LL}$, then the amplitude can be approximately reduced as
\begin{eqnarray}
&&i{\cal M}_{a1}\simeq-\bar{u}_i(p+q)
\int\frac{d^4k}{(2\pi)^D}\frac{e\mu_HM_2A_{L}B_{L}C_{L}DP_L(2p+q)^{\mu}}
{[k^2-\mu_H^2][k^2-M_2^2][k^2-m_{\tilde{\nu}_{L_j}^{R}}^2][k^2-m_{\tilde{\nu}_{L_i}^{R}}^2]}\nonumber\\&&
\times~~~\Big[\frac{k^2}{(k^2-\mu_H^2)^2}+\frac{k^2}{(k^2-M_2^2)^2}+\frac{k^2}{(k^2-\mu_H^2)(k^2-M_2^2)}
\Big]u_j(p)\epsilon_{\mu}(q)
\nonumber\\&&=i\bar{u}_i(p+q)em_{l_j}i\sigma^{\mu\nu}q_{\nu}
\frac{g_{2}^{2}}{2\Lambda^4} \frac{m_{l_i}}{m_{l_j}} \sqrt{x_{2} x_{\mu_{H}}} \Delta_{ij}^{LL}\tan \beta\Big[ k(x_{\mu_{H}}, x_{2} , x_{\tilde{\nu}_{L_j}^{R}}, x_{\tilde{\nu}_{L_i}^{R}})\nonumber\\
&&+ k(x_{2} ,x_{\mu_{H}}, x_{\tilde{\nu}_{L_j}^{R}}, x_{\tilde{\nu}_{L_i}^{R}})-f(x_{2},x_{\mu_{H}},x_{\tilde{\nu}_{L_j}^{R}}, x_{\tilde{\nu}_{L_i}^{R}})\Big]P_Lu_j(p)\epsilon_{\mu}(q).
\end{eqnarray}

As the same method with FIG.\ref{figb}(a1), the concrete amplitude of FIG.\ref{figb}(a2) can be written as
\begin{eqnarray}
&&\hspace{-2cm}i{\cal M}_{a2}\hspace{-0.1cm}=\hspace{-0.1cm}\bar{u}_i(p+q)
\hspace{-0.2cm}\int\hspace{-0.2cm}\frac{d^4k}{(2\pi)^D}\hspace{-0.1cm}\Big[\frac{eA'_RP_R({/\!\!\!k\hspace{-0.1cm}+\hspace{-0.1cm}/\!\!\!p\hspace{-0.1cm}+\hspace{-0.1cm}/\!\!\!q}\hspace{-0.1cm}+\hspace{-0.1cm}M_2)(B'_LP_L\hspace{-0.1cm}+\hspace{-0.1cm}B'_RP_R)({/\!\!\!k\hspace{-0.1cm}+\hspace{-0.1cm}/\!\!\!p\hspace{-0.1cm}+\hspace{-0.1cm}/\!\!\!q}\hspace{-0.1cm}+\hspace{-0.1cm}\mu_H)\gamma^{\mu}({/\!\!\!k\hspace{-0.1cm}+\hspace{-0.1cm}/\!\!\!p}\hspace{-0.1cm}+\hspace{-0.1cm}\mu_H)C'_RP_RD}
{[(k+p+q)^2\hspace{-0.1cm}-\hspace{-0.1cm}M_2^2][(k+p+q)^2\hspace{-0.1cm}-\hspace{-0.1cm}\mu_H^2][(k+p)^2\hspace{-0.1cm}-\hspace{-0.1cm}\mu_H^2][k^2\hspace{-0.1cm}-\hspace{-0.1cm}m_{\tilde{\nu}_{L_j}^{R}}^2][k^2\hspace{-0.1cm}-\hspace{-0.1cm}m_{\tilde{\nu}_{L_i}^{R}}^2]}
\nonumber\\&&\hspace{0cm}+\frac{eA'_RP_R({/\!\!\!k\hspace{-0.1cm}+\hspace{-0.1cm}/\!\!\!p\hspace{-0.1cm}+\hspace{-0.1cm}/\!\!\!q}\hspace{-0.1cm}+\hspace{-0.1cm}M_2)\gamma^{\mu}({/\!\!\!k\hspace{-0.1cm}+\hspace{-0.1cm}/\!\!\!p}\hspace{-0.1cm}+\hspace{-0.1cm}M_2)(B'_LP_L\hspace{-0.1cm}+\hspace{-0.1cm}B'_RP_R)({/\!\!\!k\hspace{-0.1cm}+\hspace{-0.1cm}/\!\!\!p}\hspace{-0.1cm}+\hspace{-0.1cm}\mu_H)C'_RP_RD}
{[(k+p+q)^2\hspace{-0.1cm}-\hspace{-0.1cm}M_2^2][(k+p)^2\hspace{-0.1cm}-\hspace{-0.1cm}M_2^2][(k+p)^2\hspace{-0.1cm}-\hspace{-0.1cm}\mu_H^2][k^2\hspace{-0.1cm}-\hspace{-0.1cm}m_{\tilde{\nu}_{L_j}^{R}}^2][k^2\hspace{-0.1cm}-\hspace{-0.1cm}m_{\tilde{\nu}_{L_i}^{R}}^2]}
\Big]u_j(p)\epsilon_{\mu}(q).
\end{eqnarray}
With $A'_R=\frac{g_2}{\sqrt{2}}$, $B'_L=-\frac{g_2}{\sqrt{2}}v_d$, $B'_R=-\frac{g_2}{\sqrt{2}}v_d\tan\beta$ (here $B'_L\ll B'_R$), $C'_R=\frac{Y_{l_j}}{\sqrt{2}}$ and $D=-\bigtriangleup_{ij}^{LL}$, then the amplitude can be approximately reduced as
\begin{eqnarray}
&&i{\cal M}_{a2}\simeq-\bar{u}_i(p+q)
\int\frac{d^4k}{(2\pi)^D}\frac{e\mu_HM_2A'_{R}B'_{R}C'_{R}DP_R(2p+q)^{\mu}}
{[k^2-\mu_H^2][k^2-M_2^2][k^2-m_{\tilde{\nu}_{L_j}^{R}}^2][k^2-m_{\tilde{\nu}_{L_i}^{R}}^2]}\nonumber\\&&
\times~~~\Big[\frac{k^2}{(k^2-\mu_H^2)^2}+\frac{k^2}{(k^2-M_2^2)^2}+\frac{k^2}{(k^2-\mu_H^2)(k^2-M_2^2)}
\Big]u_j(p)\epsilon_{\mu}(q)
\nonumber\\&&=i\bar{u}_i(p+q)em_{l_j}i\sigma^{\mu\nu}q_{\nu}
\frac{g_{2}^{2}}{2\Lambda^4} \sqrt{x_{2} x_{\mu_{H}}} \Delta_{ij}^{LL}\tan \beta\Big[ k(x_{\mu_{H}}, x_{2} , x_{\tilde{\nu}_{L_j}^{R}}, x_{\tilde{\nu}_{L_i}^{R}})\nonumber\\
&&+ k(x_{2} ,x_{\mu_{H}}, x_{\tilde{\nu}_{L_j}^{R}}, x_{\tilde{\nu}_{L_i}^{R}})-f(x_{2},x_{\mu_{H}},x_{\tilde{\nu}_{L_j}^{R}}, x_{\tilde{\nu}_{L_i}^{R}})\Big]P_Ru_j(p)\epsilon_{\mu}(q).
\end{eqnarray}
Therefore, in the one-loop contribution from $\tilde{H}^{-}-\tilde{W}^{-}-\tilde{\nu}_{L}^{R}$, the term including the incident lepton is dominant.


\begin{thebibliography}{90}

\vspace{3mm}

\bibitem{SMLFV}S. T. Petcov, Sov. J. Nucl. Phys., {\bf25}: 340 (1977) JINR-E2-10176

\bibitem{merexp}A. M. Baldini et al., (MEG Collaboration), Eur. Phys. J. C,
{\bf76}: 434 (2016)
\bibitem{tauerexp}B. Aubert et al., (BABAR Collaboration), Phys. Rev. Lett.,
{\bf104}: 021802 (2010)
\bibitem{taumurexp}K. Uno et al., (BELLE Collaboration), J. High Energy Phys.,
{\bf10}: 019 (2021)

\bibitem{CLFVNP1}A. Ilakovac and A. Pilaftsis, Nucl. Phys. B, {\bf437}: 491 (1995)
\bibitem{CLFVNP2}R. Diaz, R. Martinez and J. A. Rodriguez, Phys. Rev. D, {\bf63}:
095007 (2001)
\bibitem{CLFVNP3}M. Kakizaki, Y. Ogura and F. Shima, Phys. Lett. B, {\bf566}: 210
(2003)
\bibitem{CLFVNP4}E. Arganda and M. J. Herrero, Phys. Rev. D, {\bf73}: 055003
(2006)
\bibitem{CLFVNP5}T. Toma and A. Vicente, J. High Energy Phys., {\bf01}: 160 (2014)
\bibitem{CLFVNP6}H. B. Zhang, T. F. Feng, S. M. Zhao and F. Sun, Int. J. Mod.
Phys. A, {\bf29}: 1450123 (2014)
\bibitem{CLFVNP7}S. M. Zhao, T. F. Feng, H. B. Zhang et al., Phys. Rev. D, {\bf92}: 115016 (2015)
\bibitem{CLFVNP8}J. L. Yang, T. F. Feng, Y. L. Yan et al., Phys. Rev. D, {\bf99}: 015002 (2019)
\bibitem{CLFVNP9}T. Nomura, H. Okada, Y. Uesaka, Nucl. Phys. B, {\bf962}: 115236 (2021)

\bibitem{B-LSSM1}V. Barger, P. F. Perez and S. Spinner, Phys. Rev. Lett., {\bf102}: 181802 (2009)
\bibitem{B-LSSM2}P. F. Perez and S. Spinner, Phys. Lett. B, {\bf673}: 251 (2009)
\bibitem{B-LSSM3}M. Ambroso and B. A. Ovrut, Int. J. Mod. Phys. A, {\bf26}: 1569 (2011)
\bibitem{B-LSSM4}P. F. Perez and S. Spinner, Phys. Rev. D, {\bf83}: 035004 (2011)
\bibitem{B-LSSM5}J. L. Yang, S. M. Zhao, R. F. Zhu et al., Eur. Phys. J. C, {\bf78}: 714 (2018)

\bibitem{g-2MIA}T. Moroi, Phys. Rev. D, {\bf53}: 6565 (1996)
\bibitem{htaumuMIA}E. Arganda, M. J. Herrero and R. Morales et al., J. High Energy Phys.,  {\bf03}: 055 (2016)
\bibitem{HliljMIA}E. Arganda, M. J. Herrero and X. Marcano et al., Phys. Rev. D, {\bf95}: 095029 (2017)
\bibitem{ZlklmMIA}M. J. Herrero, X. Marcano and R. Morales et al., Eur. Phys. J. C, {\bf78}: 815 (2018)
\bibitem{Zhaog-2MIA}S. M. Zhao, L. H. Su, X. X. Dong et al., J. High Energy Phys.,  {\bf03}: 101 (2022)
\bibitem{WangljlirMIA}T. T. Wang, S. M. Zhao, J. F. Zhang et al., Eur. Phys. J. C, {\bf82}: 639 (2022)

\bibitem{MSSM1}H. P. Nilles, Phys. Rept., {\bf110}: 1 (1984)
\bibitem{MSSM2}H. E. Haber and G. L. Kane, Phys. Rept., {\bf117}: 75 (1985)
\bibitem{MSSM3}J. Rosiek, Phys. Rev. D, {\bf 41}: 3464 (1990)
\bibitem{MSSM4}T. F. Feng and X. Y. Yang, Nucl. Phys. B, {\bf 814}: 101 (2009)

\bibitem{B-L R Parity}C. S. Aulakh, A. Melfo, A. Rasin and G. Senjanovic, Phys. Lett. B, {\bf459}: 557 (1999)

\bibitem{B-L hierarchy1}W. Abdallah, A. Hammad, S. Khalil and S. Moretti, Phys. Rev. D, {\bf95}: 055019 (2017)
\bibitem{B-L hierarchy2}J. L. Yang, T. F. Feng and H. B. Zhang, Eur. Phys. J. C, {\bf80}: 210 (2020)

\bibitem{B-LDM1}S. Khalil and H. Okada, Phys. Rev. D, {\bf79}: 083510 (2009)
\bibitem{B-LDM2}L. Basso, B. O'Leary, W. Porod and F. Staub, J. High Energy Phys.,  {\bf1209}: 054 (2012)
\bibitem{B-LDM3}L. D. Rose, S. Khalil, S. J. D. King et al., Phys. Rev. D, {\bf96}: 055004 (2017)
\bibitem{B-LDM4}L. D. Rose, S. Khalil, S. J. D.King et al., J. High Energy Phys.,  {\bf07}: 100 (2018)

\bibitem{gBgYBB-L}P. H. Chankowski, S. Pokorski and J. Wagner, Eur. Phys. J. C, {\bf47}: 187 (2006)

\bibitem{SARAH1}F. Staub, Adv. High Energy Phys., {\bf2015}: 840780 (2015)
\bibitem{SARAH2}F. Staub, (2008) arxiv/hep-ph: 0806.0538

\bibitem{leadinglog1}M. Carena, J. R. Espinosaos, C. E. M. Wagner et al., Phys. Lett. B, {\bf355}: 209 (1995)
\bibitem{leadinglog2}M. Carena, M. Quiros and C. E. M. Wagner, Nucl. Phys. B, {\bf461}: 407 (1996)
\bibitem{leadinglog3}M. Carena, S. Gori, N. R. Shah et al., J. High Energy Phys., {\bf03}: 014 (2012)

\bibitem{effective1}T. F. Feng, L. Sun and X. Y. Yang, Nucl. Phys. B, {\bf800}: 221
(2008)
\bibitem{effective2}T. F. Feng, L. Sun and X. Y. Yang, Phys. Rev. D, {\bf77}: 116008
(2008)

\bibitem{Zpupper}G. Aad et al., (ATLAS Collaboration), Phys. Lett. B, {\bf796}: 68-87 (2019)

\bibitem{Zpupper1}G. Cacciapaglia, C. Csaki, G. Marandella et al., Phys. Rev. D, {\bf74}: 033011 (2006)
\bibitem{Zpupper2}M. Carena, A. Daleo, B. A. Dobrescu et al, Phys. Rev. D, {\bf70}: 093009 (2004)

\bibitem{BSgamma1}F. Mamoudi, J. High Energy Phys.,  {\bf12}: 026 (2007)
\bibitem{BSgamma2}K. A. Olive and L. Velasco-Sevilla, J. High Energy Phys., {\bf05}: 052 (2008)

\bibitem{gYB1}B. O'Leary, W. Porod and F. Staub, J. High Energy Phys., {\bf1205}: 042 (2012)
\bibitem{gYB2}X. X. Dong, T. F. Feng, H. B. Zhang et al., J. High Energy Phys., {\bf12}: 052 (2021)

\bibitem{PDG2022}R. L. Workman et al., (Particle Data Group), Prog. Theor. Exp. Phys., {\bf2022}: 083C01 (2022)

\bibitem{Higgsmassexp1}G. Aad et al., (ATLAS and CMS Collaborations), Phys. Rev. Lett., {\bf114}: 191803 (2015)
\bibitem{Higgsmassexp2}M. Aaboud et al., (ATLAS Collaboration), Phys. Lett. B, {\bf784}: 345 (2018)
\bibitem{Higgsmassexp3}A. M. Sirunyan et al., (CMS Collaboration), Phys. Lett. B, {\bf805}: 135425 (2020)

\bibitem{g-2exp}B. Abi et al., (Muon g-2 Collaboration), Phys. Rev. Lett., {\bf126}: 14, 141801 (2021)

\bibitem{compatibility1} A. Dedes, M. Paraskevas, J. Rosiek, et al., J. High Energy Phys., {\bf06}: 151 (2015)
\bibitem{compatibility2} J. Rosiek, Comput. Phys. Commun. {\bf201}: 144 (2016)

\bibitem{chirality flip1} J. Hisano, T. Moroi, K. Tobe, et al., Phys. Lett. B, {\bf357}: 579-587 (1995)
\bibitem{chirality flip2} R. Barbieri, L. Hall and A. Strumia, Nucl. Phys. B, {\bf445}: 219-251 (1995)
\bibitem{chirality flip3} J. Hisano, T. Moroi, K. Tobe and M. Yamaguchi, Phys. Rev. D, {\bf53}: 2442-2459 (1996)
\end{thebibliography}
 \end{document}